\documentclass{aa}  
\usepackage{graphicx}
\usepackage{txfonts}
\usepackage{natbib}
\usepackage{subcaption}
\usepackage{tablefootnote}
\usepackage[version=4]{mhchem}
\usepackage{threeparttable, tablefootnote}
\usepackage{multirow}
\usepackage{gensymb}
\usepackage{amssymb}

\begin{document} 

\title{First detections of \ce{H^{13}CO^+} and \ce{HC^{15}N} in the disk around HD 97048}
\subtitle{Evidence for a cold gas reservoir in the outer disk}
\author{Alice S. Booth, Catherine Walsh \& John D. Ilee} 
\institute{School of Physics and Astronomy, University of Leeds, Leeds LS2 9JT, UK 
\hfill \break \email{pyasb@leeds.ac.uk, c.walsh1@leeds.ac.uk}}
\date{}

\titlerunning{First detections of \ce{H^{13}CO^+} and \ce{HC^{15}N} in the disk around HD 9704}
\authorrunning{Booth, Walsh \& Ilee 2019}

\abstract{
Observations of different molecular lines in protoplanetary disks 
provide valuable information on the gas kinematics, 
as well as constraints on the radial density and temperature structure of the gas.
With ALMA we have detected \ce{H^{13}CO+} (J=4-3) and \ce{HC^{15}N} (J=4-3) in the HD~97048 protoplanetary disk for the first time.
We compare these new detections to the ringed continuum mm-dust emission and the spatially resolved 
\ce{CO} (J=3-2) and \ce{HCO+} (J=4-3) emission.
The radial distributions of the \ce{H^{13}CO+} and \ce{HC^{15}N} emission show hints of ringed sub-structure 
whereas, the optically thick tracers, \ce{CO} and \ce{HCO+}, do not.
We calculate the \ce{HCO+}/\ce{H^{13}CO+} intensity ratio across the disk and find that it is radially constant (within our uncertainties). 
We use a physio-chemical parametric disk structure of the HD~97048 disk with an analytical prescription for the \ce{HCO+} abundance distribution 
to generate synthetic observations of the \ce{HCO+} and \ce{H^{13}CO+} disk emission assuming LTE.
The best by-eye fit models require radial variations in the \ce{HCO+}/\ce{H^{13}CO+} abundance ratio
and an overall enhancement in \ce{H^{13}CO+} relative to \ce{HCO+}.
This highlights the need to consider isotope selective chemistry and in particular low temperature carbon isotope exchange reactions. 
This also points to the presence of a reservoir of cold molecular gas in the outer disk (T~$\lesssim$~10~K, R~$\gtrsim$~200~au).
Chemical models are required to confirm that isotope-selective chemistry alone can explain the observations presented here. 
With these data, we cannot rule out that the known dust substructure in the HD~97048 disk is responsible for the
observed trends in molecular line emission, and higher spatial resolution observations are required
to fully explore the potential of optically thin tracers to probe planet-carved dust gaps. 
We also report non-detections of \ce{H^{13}CO+} and \ce{HC^{15}N} in the HD~100546 protoplanetary disk.}

\maketitle

\section{Introduction}

Spatially resolved Atacama Large Millimeter/submillimeter Array (ALMA) observations of the (sub-)millimetre-sized dust grains in 
protoplanetary disks are providing incredible insights into the processes 
controlling how disks evolve and how planets form. 
ALMA has detected multiple concentric rings of continuum emission
from disks around both Herbig Ae/Be and T Tauri stars,
e.g., HD~100546 and HL~Tau \citep{2014ApJ...791L...6W, 2015ApJ...808L...3A}.
Disks around Herbig Ae stars are the progenitors of gas-giant planetary systems around A-type stars, 
which have the highest occurrence rate of gas-giant planets across the stellar mass range \citep[][]{2015A&A...574A.116R, 2019AJ....158...13N}.
The observed structures in disks have been proposed to 
signify planet formation where the apparent gaps are the result of
dynamical clearing due to forming planets or massive companions \citep[e.g.,][]{2012A&A...545A..81P}. 
Recent results show that rings are the most common 
sub-structure observed in the mm-sized dust and the lack of a clear trend in 
these structures with host star properties favours planets as the likely explanation \citep{2018ApJ...869L..41A, 2018ApJ...869L..42H}.

Gas cavities in the inner disk associated with dust cavities 
have been resolved in \ce{CO} isotopologue emission and can be used to determine the cause 
of the cavity \citep{2015A&A...579A.106V, 2016A&A...585A..58V}. 
Gas density perturbations associated with dust rings $\sim 100$'s of au from the central star are less commonly observed. 
Examples have been seen with ALMA in the HD~163296, HD~169142 and AS 209 protoplanetary disks (\citealt{2016PhRvL.117y1101I}, \citealt{2017A&A...600A..72F}, \citealt{2019ApJ...871..107F})
where the \ce{^{12}CO}, \ce{^{13}CO} and \ce{C^{18}O} line observations do not 
follow the exact same radial profiles as the dust, but there is a
change in slope of these line intensity profiles that coincides with the location 
of the dark rings postulated to be dust gaps.

Although \ce{CO} isotopologues are the most common tracers of the molecular gas in disks,  
it is beneficial to observe lines from other molecules to gain more information about the physical and chemical conditions of the gas. 
Since gas-phase \ce{CO} is present in high abundance throughout the most of the disk, 
observations of \ce{CO} can suffer from optical depth effects 
such that the emission traces the upper disk layers only.
Observations of increasingly rarer isotopologues, e.g., \ce{^{13}C^{18}O} \citep[][]{2017NatAs...1E.130Z} and \ce{^{13}C^{17}O} \citep[][]{booth2019a}, can help to mitigate these optical depth effects.
Species other than \ce{CO} can be more powerful tracers of physical conditions because the abundances are more sensitive to changes in the disk's physical conditions.
For most of the disk, the chemistry of \ce{CO} can be described simply by freeze out and photodissociation,
whereas other species' abundances are more dependent on the, e.g., gas number density and ionisation fraction.

A legacy of observations of protoplanetary disks with single dish telescopes have showed that \ce{HCO+} and \ce{HCN} are two of the brightest molecules in these objects
\citep[e.g.,][]{1997Sci...277...67K, 1997A&A...317L..55D, 2001A&A...377..566V, 2004A&A...425..955T, 2007A&A...467..163P, 2012A&A...537A..60C}. 
Emission from \ce{HCO+} and \ce{HCN} and selected isotopologues have now been spatially resolved 
in a number of disks with 
ALMA \citep[e.g.][]{2017ApJ...835..231H, 2017ApJ...836...30G}
and exhibit centrally peaked or ringed radial emission profiles. 
These observed morphologies highlight the different physical and chemical processes that shape the abundance distributions of molecules in disks.

\ce{HCO+} emission from the main isotopologue tends to be optically thick in disks and traces the dense gas. 
From the aforementioned observations and supporting chemical models \citep[e.g.,][]{2015ApJ...807..120A} this molecular cation resides primarily in the warm molecular layer. 
\ce{HCO+} is commonly used to investigate disk ionisation and kinematics 
\citep[e.g.,][]{1999A&A...348..570G, 2001A&A...377..566V, 2010ApJ...720..480O, 2011ApJ...734...98O, 2014ApJ...794..123C, 2015A&A...574A.137T}. 
The chemical models conducted by \citet{2015ApJ...799..204C} coupled with observations,
including of \ce{HCO+}, reported evidence for the exclusion of cosmic rays 
from a protoplanetary disk due to stellar winds and/or the magnetic field of the accreting host star. 
Spatially unresolved observations of \ce{HCO+} emission from LkCa~15 with JCMT 
have been used to constrain the gas-to-dust mass ratio in the dust cavity by modelling the 
line wing flux \citep{2016ApJ...833..260D}. 
They find that in the cavity the gas-to-dust mass ratio has to be increased by $\gtrsim 10^{4}$ relative to the ISM 
indicating a significant reservoir of molecular gas within the dust cavity.
Deviations from global Keplerian gas motion in the inner cavities of disks
have also been inferred through inspection of the intensity weighted-velocity field 
of \ce{HCO+} line emission, i.e., the first-moment map
\citep{2013Natur.493..191C, 2014ApJ...782...62R, 2015ApJ...811...92C,2017ApJ...840...23L}.

Another abundant disk molecule is \ce{HCN}. The \ce{CN}$/$\ce{HCN} ratio is used to trace the
UV radiation field across disks because
\ce{HCN} can only exist in regions where the surface density is high 
enough to attenuate this radiation, thus preventing photodissociation \citep{2010ApJ...720..480O, 2012A&A...537A..60C}. 
In comparison, \ce{CN} emission is observed to be more radially 
extended, especially in Herbig disks, because it can survive in the low density atmosphere.
Rings of \ce{CN} emission have been detected in a few protoplanetary disks with ALMA 
\citep[e.g., TW Hydrae;][]{2016A&A...592A..49T}.
Chemical modelling shows that these rings arise naturally in disks as a result of chemistry 
and are not formed as a result of an underlying ringed dust structure \citep{2018A&A...609A..93C}.
The isotopologue, \ce{HC^{15}N}, was first detected in the MWC~480 Herbig Ae protoplanetary disk 
\citep{2015ApJ...814...53G}.
The \ce{^{14}N}$/$\ce{^{15}N} ratio observed in protoplanetary disks can help 
us to better understand the origin of the nitrogen fractionation and thus chemical history of our solar system \citep{2015NatGe...8..515F}.
In the MWC~480 disk the \ce{^{14}N}$/$\ce{^{15}N} ratio determined from observations of 
CN, HCN, \ce{H^{13}CN, and \ce{HC^{15}N}} 
is 200$\pm$100 and this is similar to the value derived from \ce{C^{14}N} and \ce{C^{15}N} observations of TW Hya \citep[323$\pm$32][]{2017A&A...603L...6H}.
These values are low
relative to solar wind value of $\approx$~440 \citep{Marty1533}.
The solar wind value is similar to what is measured for cloud cores and comets 
indicating active fractionation processes occurring in both cores and disks that are then 
preserved in the cometary record.

From the observations conducted thus far, it is clear that 
molecular line observations from tracers such as \ce{HCO+} and HCN in protoplanetary disks 
provide valuable information on the physical conditions and kinematics of the emitting gas, 
as well as constraints on the radial structure of the gas.
Here we present the first detections of \ce{H^{13}CO+} and \ce{HC^{15}N} in the 
disk around the Herbig Ae star, HD~97048 using ALMA Cycle 0 observations.
Using our \ce{CO} observations and complementary archival 
ALMA Cycle 2 observations of \ce{HCO+}, we compare the radial distribution of these four
different gas tracers to the mm-sized dust emission. 
The aim of this work is to investigate whether or not 
chemistry alone can explain the radial emission profiles of these molecules, 
or if modifications to the underlying disk gas structure are required.
Section 2 gives an overview of previous observations of the source, 
Section 3 details the observations and data reduction,  
Section 4 presents the molecular line detections. 
We then describe our modelling in Section 5, in Section 6 we discuss our results,  
and in Section 7 we state our conclusions. 

\section{The source: HD~97048}

HD~97048 is a 2.5 M$_{\odot}$ Herbig Ae/Be star with spectral type B9/A0 \citep{1998A&A...330..145V, 2007A&A...474..653V} 
located in the Chameleon I star-forming region at 11:08:03.2, -77:39:17 at a distance of $184^{+2}_{-2}$~pc \citep{2018A&A...620A.128V}. 
This young ($\approx~4$~Myr) star is host to a large circumstellar disk seen with 
an inclination angle of $41 \degree$ 
and a position angle of $3 \degree$ \citep[e.g.,][]{2016ApJ...831..200W}.
We note that the disk sizes and radial positions that follow have been scaled to account for the 
new source distance as determined from GAIA Data Release 2 \citep{2018A&A...616A...1G}.  

HD~97048 is a source that has been well studied from optical to infra-red (IR) wavelengths, 
with observations revealing much substructure in the warm gas and small dust grains. 
The transitional nature of this disk was first revealed by \citet{2013A&A...555A..64M} from modelling
the spectral energy distribution (SED) and azimuthally averaged mid-IR brightness profile. 
The best fit model for these observations is 
a dust cavity out to a radius of 40$\pm$5~au and an inner dust disk between 0.4 and 3~au. 
The detection of [OI] line emission also suggests the presence of an inner gas disk \citep{Acke2006}. 
Spectrally resolved observations of CO ro-vibrational transitions 
show that the molecular gas extends to at least 102 au and that there is a cavity in the CO emission 
out to 11~au \citep{VanderPlas2009, 2015A&A...574A..75V}. 
The origin of the CO cavity is proposed to be due to photodissociation of the CO gas by stellar radiation.

The first resolved near-IR scattered light images of the HD~97048 disk revealed the inner 19 to 190~au and 
alluded to a gap in the disk from 110 to 150~au \citep{2012A&A...538A..92Q}.
Subsequent high-resolution observations with VLT/SPHERE show four rings in the 
disk with observations covering 45 to 400~au from the star \citep{2016A&A...595A.112G}.
The disk has also been observed with the Hubble Space Telescope where circumstellar material 
was observed between 365 au and 736 au from the star \citep{2007AJ....133.2122D}.

The first spatially resolved observations of the mm-sized dust and molecular gas were 
presented by \citet{2016ApJ...831..200W} using ALMA and these data revealed the 
full spatial extent of the molecular disk. 
The mm-sized dust is observed to extend to 400~au in radius; 
while, in comparison, the flared molecular disk, revealed in CO $J=3-2$ emission, is significantly larger and extends to 860~au. 
These observations also support the transitional and ringed nature of the disk. 
There is a decrease in the continuum flux within 55~au and there are three continuum emission rings peaking 
at 55, 160 and 290~au. 
More recent ALMA observations have now resolved the inner cavity at mm wavelengths \citep{2017A&A...597A..32V}.
These latter observations also report the detection of emission from the \ce{HCO+} $J=4-3$ transition as well as the \ce{CO} $J=3-2$ transition. 
The \ce{HCO+} emission is more compact than that for CO, extending out to 540~au only. 
There is also evidence for deviations from global Keplerian rotation in the \ce{HCO+} emission on small scales, 
possibly due to radial flows or an inclined inner disk \citep{2017A&A...597A..32V}.

Although this is the brightest disk observed at sub-mm wavelengths, it remains poorly studied in terms of its molecular content.
Particularly lacking are spatially resolved observations of key molecular tracers aside from those reported above.
Nevertheless, thanks in part to observations with {\em Herschel}, it is known to host a chemically-rich gaseous disk. 
Spatially unresolved emission lines that have been observed include those from 
\ce{H2}, [OI], [CII], OH and \ce{CH+} \citep{2011A&A...533A..39C, 2012A&A...544A..78M, 2013A&A...559A..77F, 2013A&A...559A..84M}.

\section{Observations}
\label{observations}

\begin{table*}
\centering
\begin{threeparttable}
\caption{ALMA Band 7 observations and detected molecular lines in the HD~97048 protoplanetary disk. \label{table1}}
\begin{tabular}{ccccccc}
\hline\hline
\multicolumn{1}{l}{Observations}                             & \multicolumn{3}{c}{Cycle 0 (2011.0.00863.S)}     & \multicolumn{1}{c}{~~~~~}	& \multicolumn{2}{c}{Cycle 2 (2013.1.00658.S)} \\
\multicolumn{1}{l}{Date observed}                            & \multicolumn{3}{c}{14 December 2012}              & \multicolumn{1}{c}{~~~~~}& \multicolumn{2}{c}{22 May 2015} \\
\multicolumn{1}{l}{Baselines (m)}                            & \multicolumn{3}{c}{15.1 - 402}                    & \multicolumn{1}{c}{~~~~~}& \multicolumn{2}{c}{21.4 - 555.5} \\ 
\multicolumn{1}{l}{On source integration time (min,sec)}     & \multicolumn{3}{c}{23,59}                         & \multicolumn{1}{c}{~~~~~}& \multicolumn{2}{c}{7,10} \\ 
\multicolumn{1}{l}{Number of antenna}     					 & \multicolumn{3}{c}{22}                         & \multicolumn{1}{c}{~~~~~}& \multicolumn{2}{c}{36} \\ 
\multicolumn{1}{l}{CLEAN image weighting}                    & \multicolumn{6}{c}{natural} \\
\hline
\multicolumn{1}{l}{Molecule}                                 & \multicolumn{1}{c}{CO}                           & \multicolumn{1}{c}{\ce{H^{13}CO+}}               & \multicolumn{1}{c}{\ce{HC^{15}N}}	               & \multicolumn{1}{c}{~~~~~}& \multicolumn{1}{c}{CO}                           & \multicolumn{1}{c}{\ce{HCO+}} \\
\multicolumn{1}{l}{Transition}                               & \multicolumn{1}{c}{$J=3-2$}                      & \multicolumn{1}{c}{$J=4-3$}                      & \multicolumn{1}{c}{$J=4-3$}                       & \multicolumn{1}{c}{~~~~~}& \multicolumn{1}{c}{$J=3-2$}                      & \multicolumn{1}{c}{$J=4-3$} \\
\multicolumn{1}{l}{Frequency (GHz)}                          & \multicolumn{1}{c}{345.796}                      & \multicolumn{1}{c}{346.998}                      &\multicolumn{1}{c}{344.200}                        & \multicolumn{1}{c}{~~~~~}& \multicolumn{1}{c}{345.796}                      & \multicolumn{1}{c}{356.734} \\
\multicolumn{1}{l}{Einstein A coefficient (s$^{-1}$)}        & \multicolumn{1}{c}{2.497e-06}                    & \multicolumn{1}{c}{3.288e-03}                   &\multicolumn{1}{c}{1.879e-03}                     & \multicolumn{1}{c}{~~~~~}& \multicolumn{1}{c}{2.497e-06}                    & \multicolumn{1}{c}{3.627e-03} \\
\multicolumn{1}{l}{$E_\mathrm{up}$(K)}                       & \multicolumn{1}{c}{33.19}                        & \multicolumn{1}{c}{41.63}                        & \multicolumn{1}{c}{41.30}                         & \multicolumn{1}{c}{~~~~~}& \multicolumn{1}{c}{33.19}                        & \multicolumn{1}{c}{42.80} \\
\multicolumn{1}{l}{Synthesised beam}                         & \multicolumn{1}{c}{0\farcs74~$\times$~0\farcs51} & \multicolumn{1}{c}{0\farcs74~$\times$~0\farcs51} & \multicolumn{1}{c}{0\farcs74~$\times$~0\farcs52}  & \multicolumn{1}{c}{~~~~~}& \multicolumn{1}{c}{0\farcs71~$\times$~0\farcs44} & \multicolumn{1}{c}{0\farcs67~$\times$~0\farcs42} \\
\multicolumn{1}{l}{Beam P.A.}			                     & \multicolumn{1}{c}{-26\degree}                   & \multicolumn{1}{c}{-26\degree}                   & \multicolumn{1}{c}{-27\degree}                    & \multicolumn{1}{c}{~~~~~}& \multicolumn{1}{c}{-161\degree}                  & \multicolumn{1}{c}{-161\degree} \\
\multicolumn{1}{l}{Spectral resolution (km~s$^{-1}$)}        & \multicolumn{1}{c}{0.21}                         & \multicolumn{1}{c}{0.21}                         & \multicolumn{1}{c}{0.21} 	                       & \multicolumn{1}{c}{~~~~~}& \multicolumn{1}{c}{0.10}                         & \multicolumn{1}{c}{0.10} \\
\multicolumn{1}{l}{$\sigma$ (Jy~beam$^{-1}$ per channel$^{a}$)} & \multicolumn{1}{c}{0.016}                  & \multicolumn{1}{c}{0.016}                        & \multicolumn{1}{c}{0.016}                         & \multicolumn{1}{c}{~~~~~}& \multicolumn{1}{c}{0.016}                        & \multicolumn{1}{c}{0.019} \\
\hline
\end{tabular}
\begin{tablenotes}\footnotesize
\item{The values for the line frequencies, Einstein A coefficients, and upper energy levels ($E_{up}$) 
are from the Leiden Atomic and Molecular Database: \url{http://home.strw.leidenuniv.nl/~moldata/} \citep[LAMDA;][]{2005A&A...432..369S}.}
\item{$^{a}$ At the stated spectral resolution.}
\end{tablenotes}
\end{threeparttable}
\end{table*}

\subsection{Data reduction}
\label{datareduction}

Table~\ref{table1} lists the two sets of ALMA observations used in this work as well as the detected molecular lines.
The continuum emission, and \ce{CO} and \ce{HCO+} line emission from the two datasets have already 
been presented in previous work (see \citealt{2016ApJ...831..200W} and \citealt{2017A&A...597A..32V} for full details).
This work uses the self-calibrated, phase-corrected, and continuum-subtracted measurement sets from the 
Cycle 0 program 2011.0.00863.S (P.I.~C.~Walsh) and 
the raw archival data from the Cycle 2 program 2013.1.00658.S (P.I.~G.~van der Plas). 
These latter data were self-calibrated using CASA (version 4.6.0).
This dataset consists of four spectral windows, two of which were tuned for continuum observations 
and two of which were centred on the \ce{CO} $J=3-2$ and \ce{HCO+} $J=4-3$ spectral lines.
The two broadband continuum windows were self-calibrated after flagging the line containing channels and 
then used to self-calibrate the respective higher spectral resolution \ce{CO} $J=3-2$ 
and \ce{HCO+} $J=4-3$ line containing spectral windows.
We ensured that the data was phase-centred on source and subtracted the continuum emission from the two line 
containing spectral windows using the CASA task, \texttt{uvcontsub}.
Additionally, we checked that continuum subtraction in the $uv$ domain did not 
over-subtract the \ce{CO} or \ce{HCO+} emission 
\citep{2017ApJ...840...60B}. 
This was done by comparing two line profiles extracted from data cubes with and without continuum subtraction in the $uv$ domain. 
In the latter case, a simple baseline continuum subtraction was done directly on the extracted line profile.
It was found that the line profiles for both species (CO and \ce{HCO+}) using both methods agree to within $1\sigma$,
where $\sigma$ is the root-mean-square (rms) noise as determined from the nearby line-free channels. 
 
\subsection{Molecular line imaging}

The lines listed in Table~\ref{table1} 
were imaged in CASA using CLEAN with natural weighting and using  
a Keplerian mask\footnote{Publicly available on GitHub at \url{https://github.com/kevin-flaherty/ALMA-Disk-Code/blob/master/makemask.py}}.
The use of such a mask generates improved moment maps \citep[e.g.][]{2017A&A...606A.125S}.
The Keplerian mask was set to encompass emission from 1~au to 800~au using 
an estimated local linewidth of 1~km~s$^{-1}$ at 100~au.
Table~\ref{table1} lists the rms noise ($\sigma$) of each line 
imaged at its respective Hanning-smoothed spectral resolution, as well as the image parameters, 
and the intrinsic properties of the transitions. 
A comparison of the peak flux densities and integrated intensities for the \ce{CO} $J=3-2$ line that was observed in both datasets 
used here shows that they agree to within 10\% \citep[see][for details]{2016ApJ...831..200W}. 

The \ce{CO} $J=3-2$ line emission suffers from missing emission about 
the source velocity, from 3.4 to 5.7~km~s$^{-1}$ inclusive. 
This is the result of either spatial filtering due to bright background emission or foreground absorption. 
Spatial filtering is a possibility because the source is in proximity to the Chameleon I molecular cloud. 
This effect is seen in both the channel maps and 
line profiles presented by \citet{2016ApJ...831..200W} and \citet{2017A&A...597A..32V}.
This work focuses on the detection of \ce{H^{13}CO+} and \ce{HC^{15}N}
and there is no indication that they are also affected by this phenomenon. 
A possible explanation is that this is due to the higher-lying energy levels of these transitions (see Table~\ref{table1}) 
or, due to the lower abundances of \ce{HCO+}, \ce{H^{13}CO+}, and \ce{HC^{15}N}, in either background or foreground 
material, when compared with \ce{CO}. 

\subsection{Non-detections in the HD~100546 protoplanetary disk}

The protoplanetary disks around HD~97048 and HD~100546 were observed with the same 
spectral setup during the Cycle 0 ALMA campaign described above (2011.0.00863.S).  
We did not detect the \ce{H^{13}CO+} $J=4-3$ and \ce{HC^{15}N} $J=4-3$ transitions 
in our complementary observations of HD~100546 \citep[see][]{2014ApJ...791L...6W, 2018A&A...611A..16B}.
The lines were imaged in CLEAN using natural weighting and the $1\sigma$ rms noise level in the 
channel maps at a velocity resolution of 1~km~s$^{-1}$ is 8~mJy~beam$^{-1}$
per channel and 10~mJy~beam$^{-1}$ per channel for \ce{H^{13}CO+} and \ce{HC^{15}N}, respectively.
This result highlights the differences between the molecular nature of these two Group~I 
Herbig Ae/Be disks (as also discussed for the case of SO by \citealt{2018A&A...611A..16B}). 
We will further discuss these implications in Section 6. 

\section{Results}
\label{results}

\subsection{Channel maps, moment maps, and line profiles}

\begin{figure*}
\includegraphics[trim={1cm 1cm 0 4cm},clip,width=\hsize]{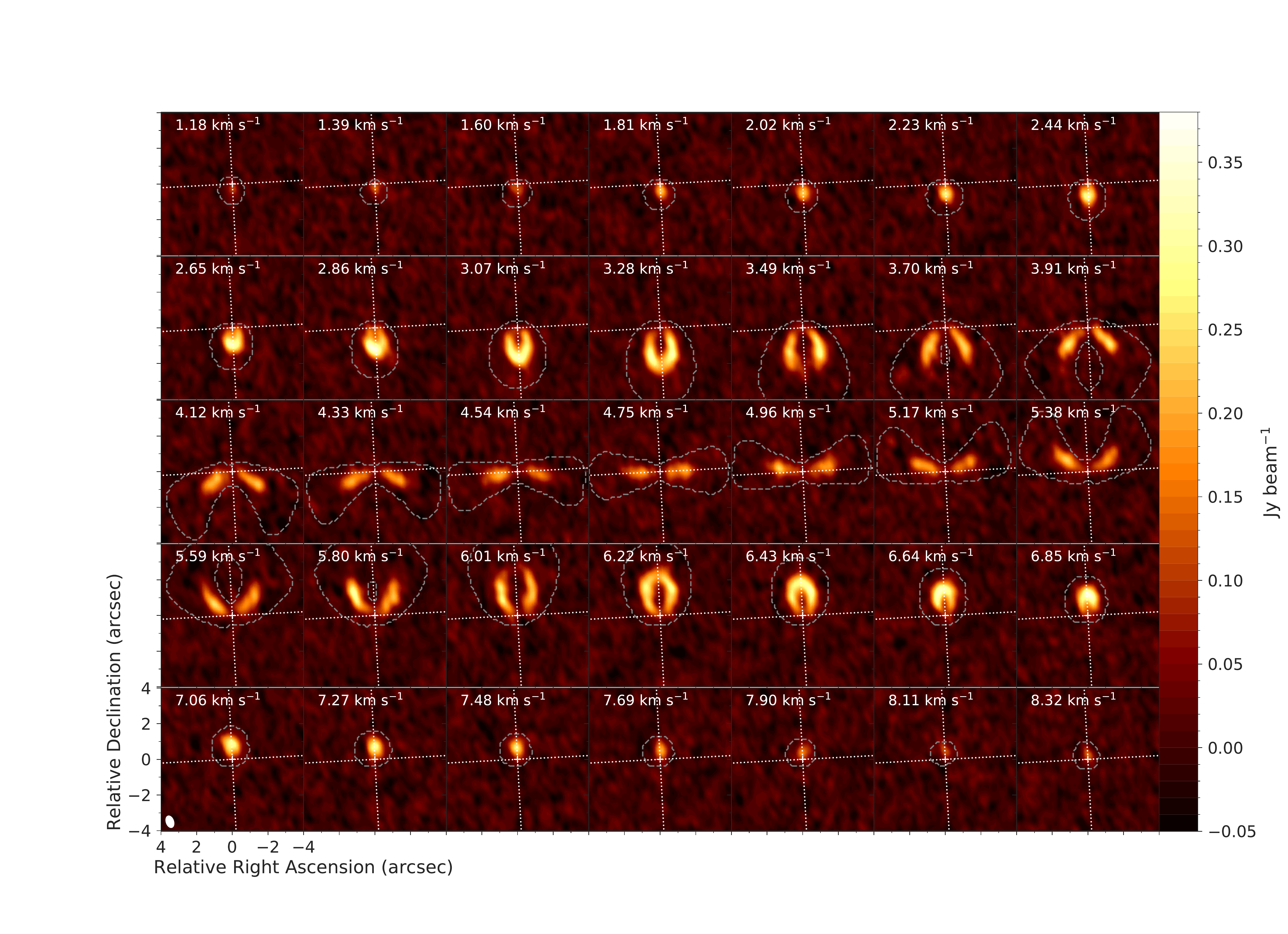}
\includegraphics[trim={1cm 1cm 0 4cm},clip,width=\hsize]{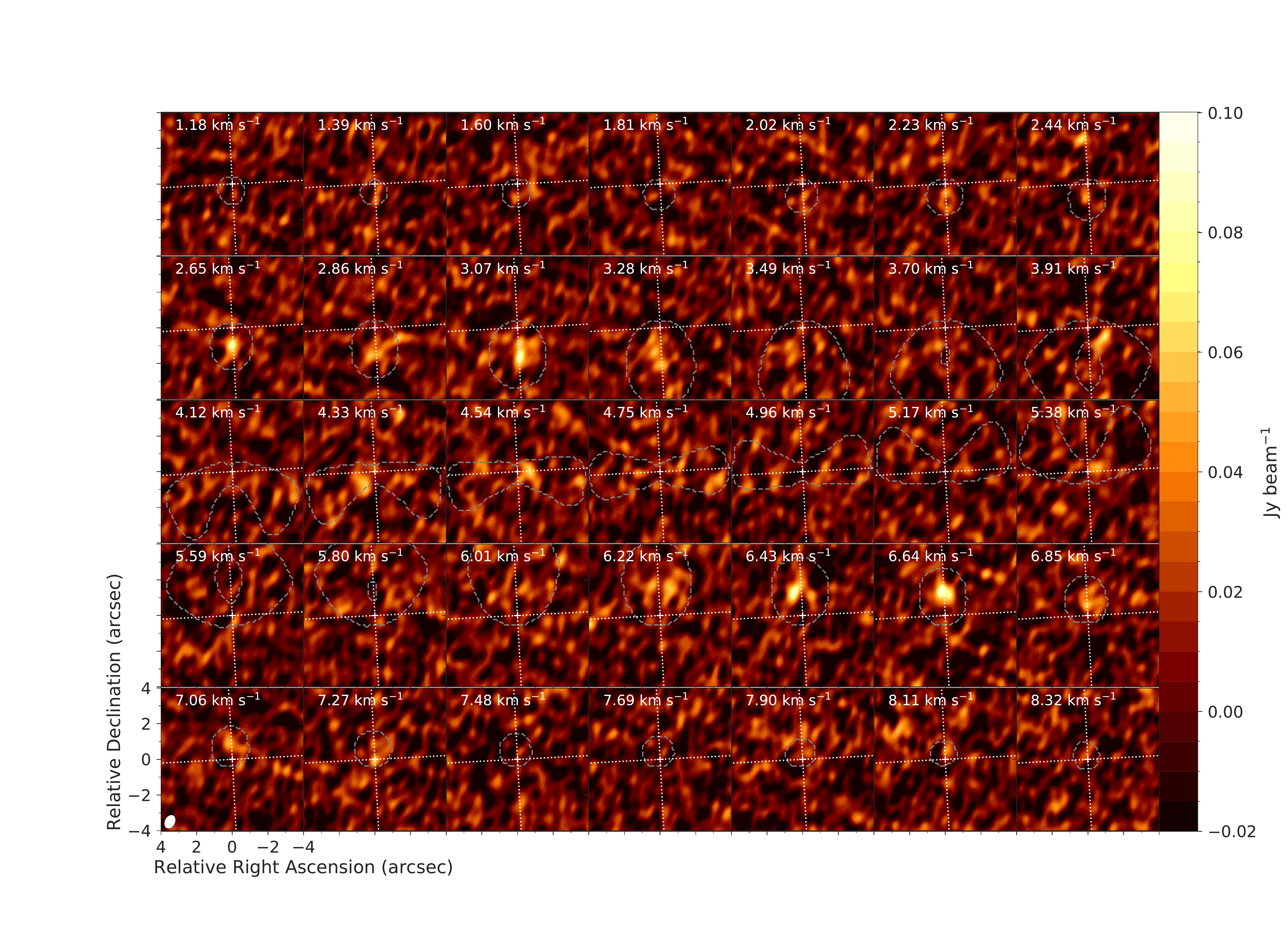}
\caption{\ce{HCO+} (top) and \ce{H^{13}CO+} (bottom) channel maps with a velocity resolution of 0.21~km~s$^{-1}$. 
The rms noise, $\sigma$, is 0.016~Jy~beam$^{-1}$ per channel in both maps and the emission is detected with a S/N of 25 and 7 respectively.
The grey dashed contour shows the Keplerian mask used and
the dotted white lines mark the major and minor axes of the disk.}
\label{figure1}
\end{figure*}

Figure~\ref{figure1} shows the \ce{HCO+} and \ce{H^{13}CO+} channel maps at 
the native spectral resolution of the \ce{H^{13}CO+} data, 0.21~km~s$^{-1}$.
We slightly over-sample the \ce{HCO+} data in order to allow a direct comparison between the two molecules.
Both channel maps have a $1\sigma$ rms noise level of 0.016~Jy~beam$^{-1}$~channel$^{-1}$ 
and reach a S/N of 23 and 6, respectively. 
To obtain a robust detection of \ce{HC^{15}N} in the channel maps 
the data were re-binned to a velocity resolution of 0.5~km~s$^{-1}$, 
and these channel maps are shown in Appendix A. 
The \ce{H^{13}CO+} channel maps at the same velocity resolution are also shown in Appendix B for ease of comparison.

For \ce{HCO+}, the highest velocities at which emission is robustly detected ($>3\sigma$ within a 0\farcs5 aperture) are in channels 
at -0.17~km~s$^{-1}$ and 9.55~km~s$^{-1}$, 
for the blue- and red-shifted emission, respectively. 
Assuming a disk inclination of 41\degree and a stellar mass of 2.5 M$_{\odot}$, 
\ce{HCO+} emission is detected down to an average radial distance of $41\pm2$~au.
This is different to the value of 28~au stated by \citet{2017A&A...597A..32V}; however, 
note that we use different imaging parameters.  
Here, we adopt natural weighting in the imaging to enhance the S/N for the detected lines. 
On the other hand, \citet{2017A&A...597A..32V} used Briggs weighting in their imaging resulting in a smaller synthesised beam.
In comparison, the highest velocity CO emission from both the Cycle 0 and Cycle 2 data as determined here
corresponds to emission detected down to an average radial distance of $14\pm1$~au.  
This is in agreement with \citet{2017A&A...597A..32V}.
The errors in these radii are propagated from the velocity resolution of the channel maps (see Table~\ref{table1}). 

\begin{figure*}
\includegraphics[trim={0cm 1.5cm 0cm 2.5cm},clip,width=0.5\hsize]{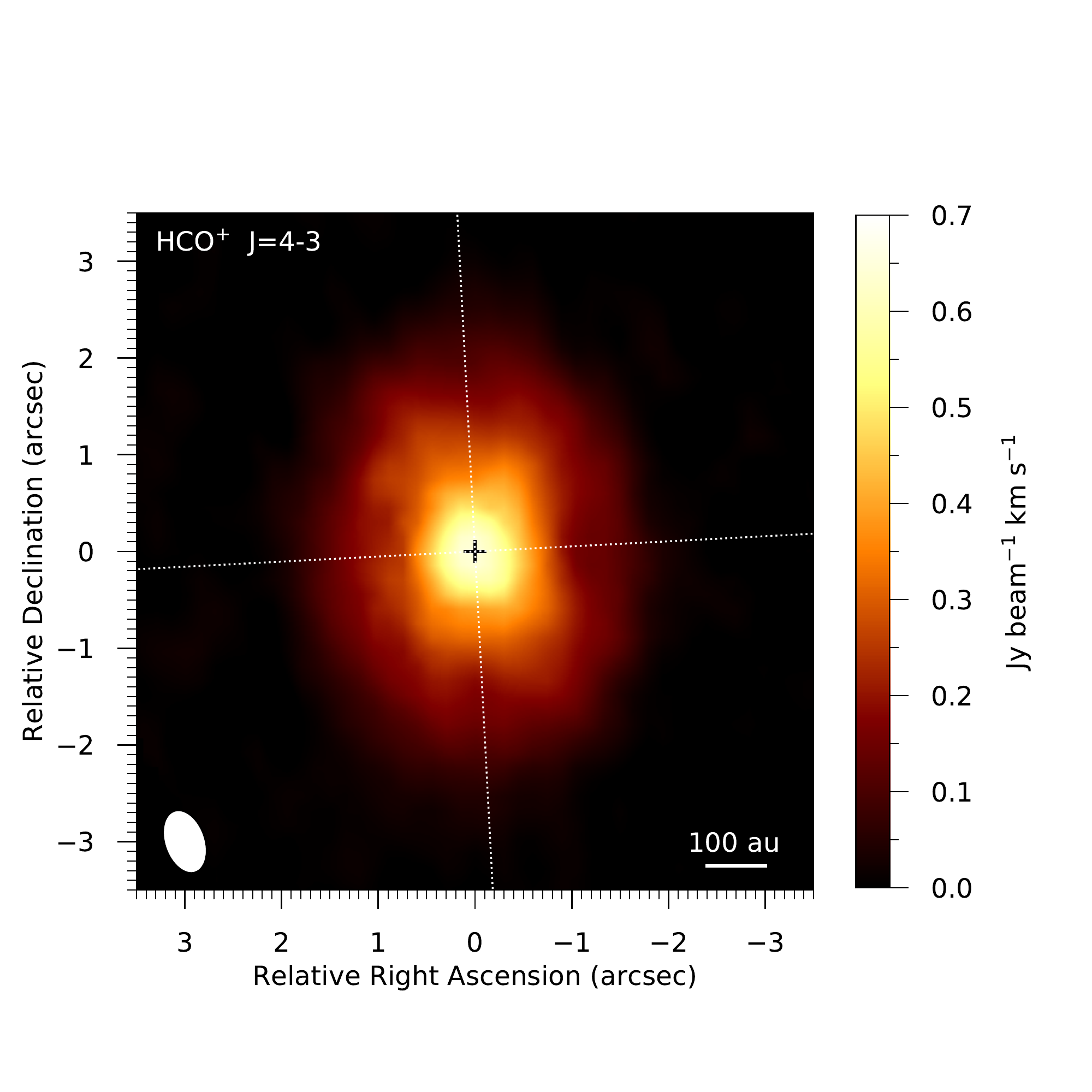}
\includegraphics[trim={0cm 1.5cm 0cm 2.5cm},clip,width=0.5\hsize]{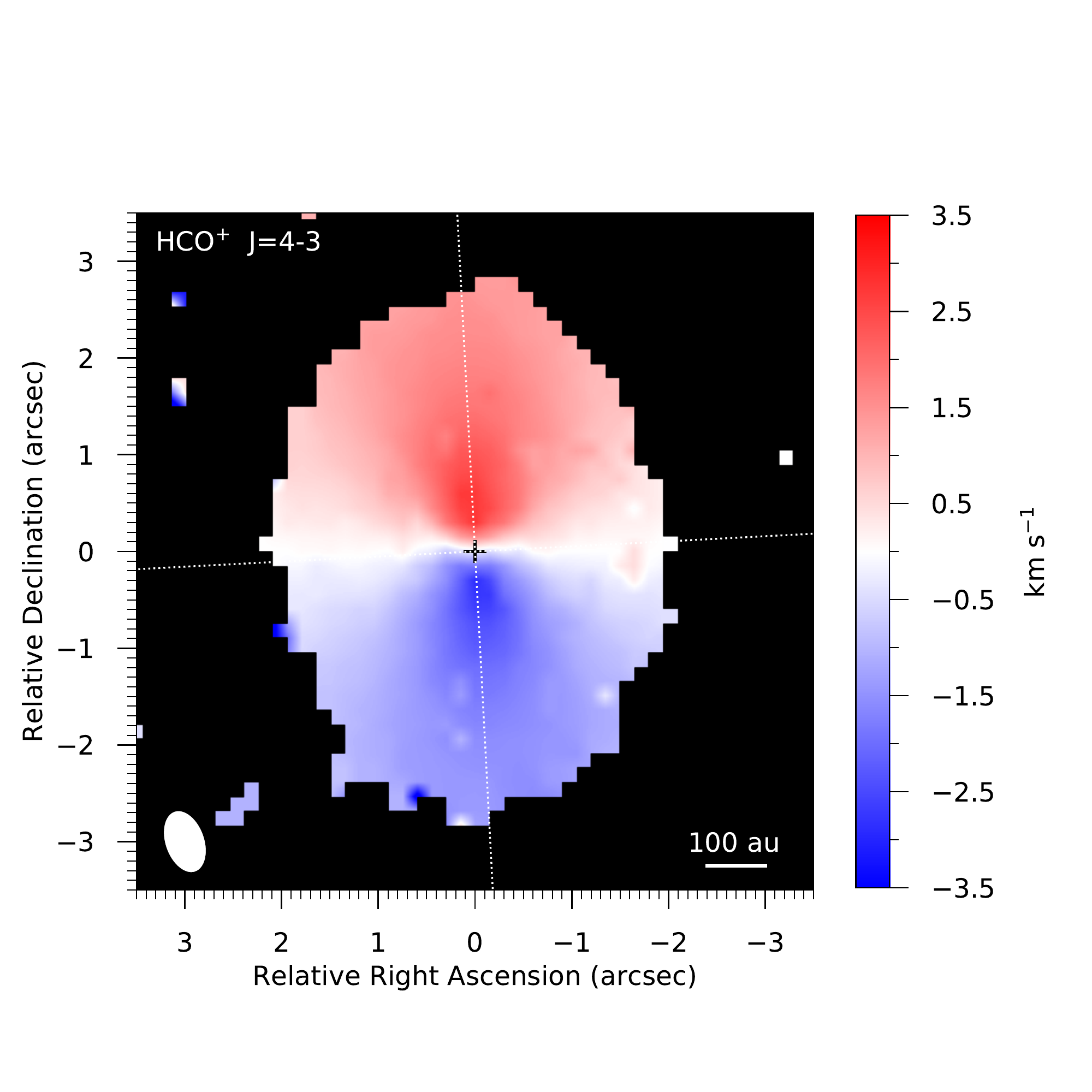}
\includegraphics[trim={0cm 1.5cm 0cm 2.5cm},clip,width=0.5\hsize]{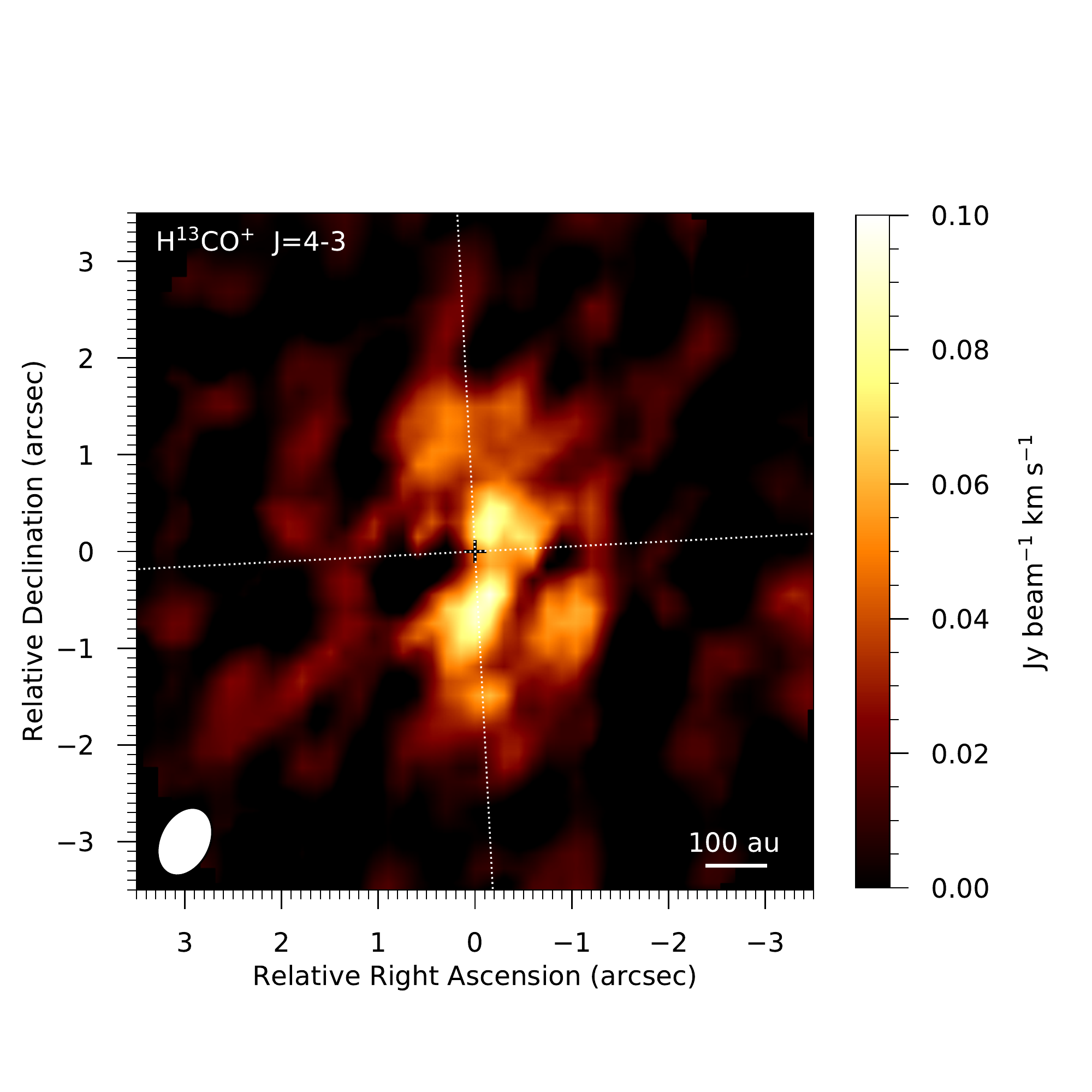}
\includegraphics[trim={0cm 1.5cm 0cm 2.5cm},clip,width=0.5\hsize]{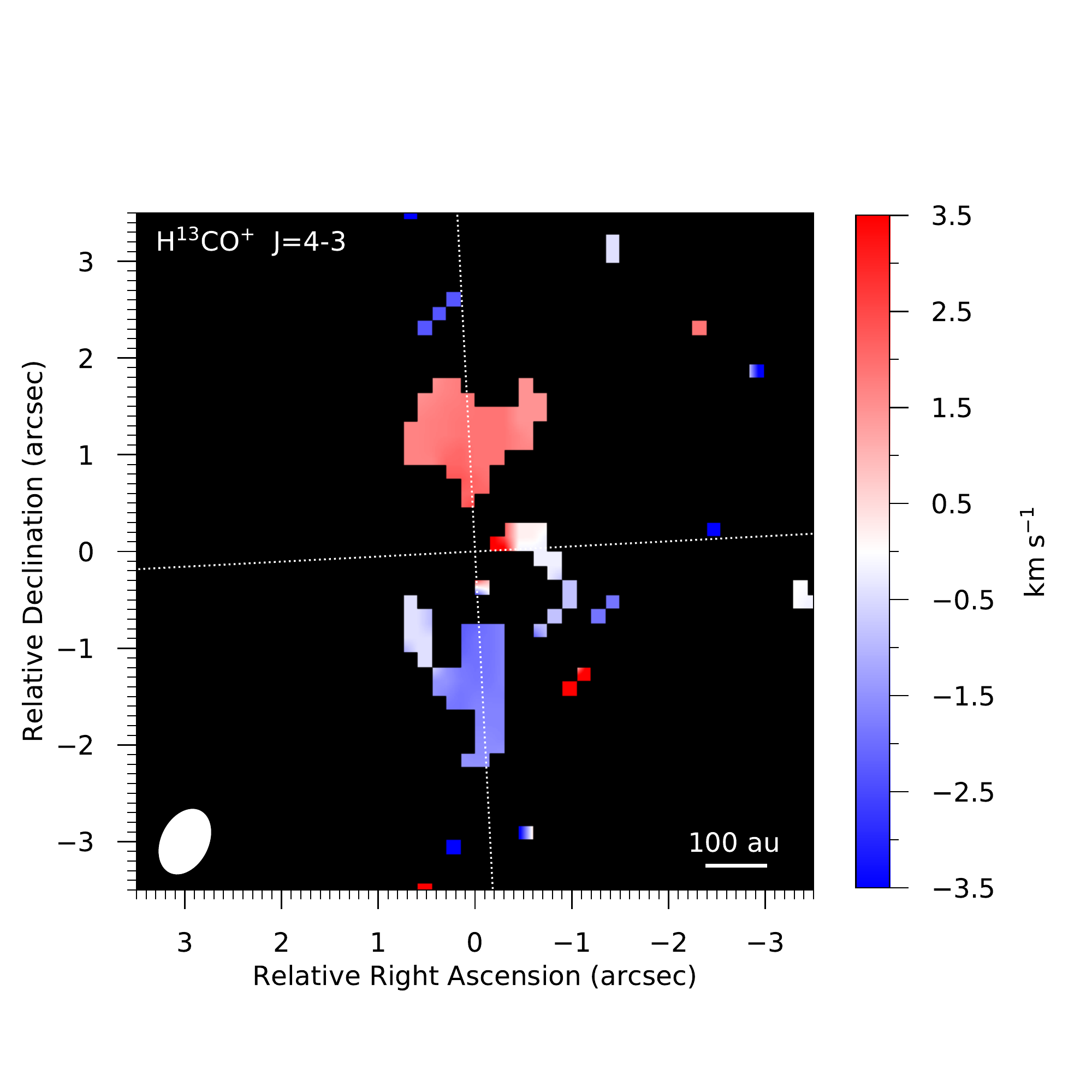}
\includegraphics[trim={0cm 1.5cm 0cm 2.5cm},clip,width=0.5\hsize]{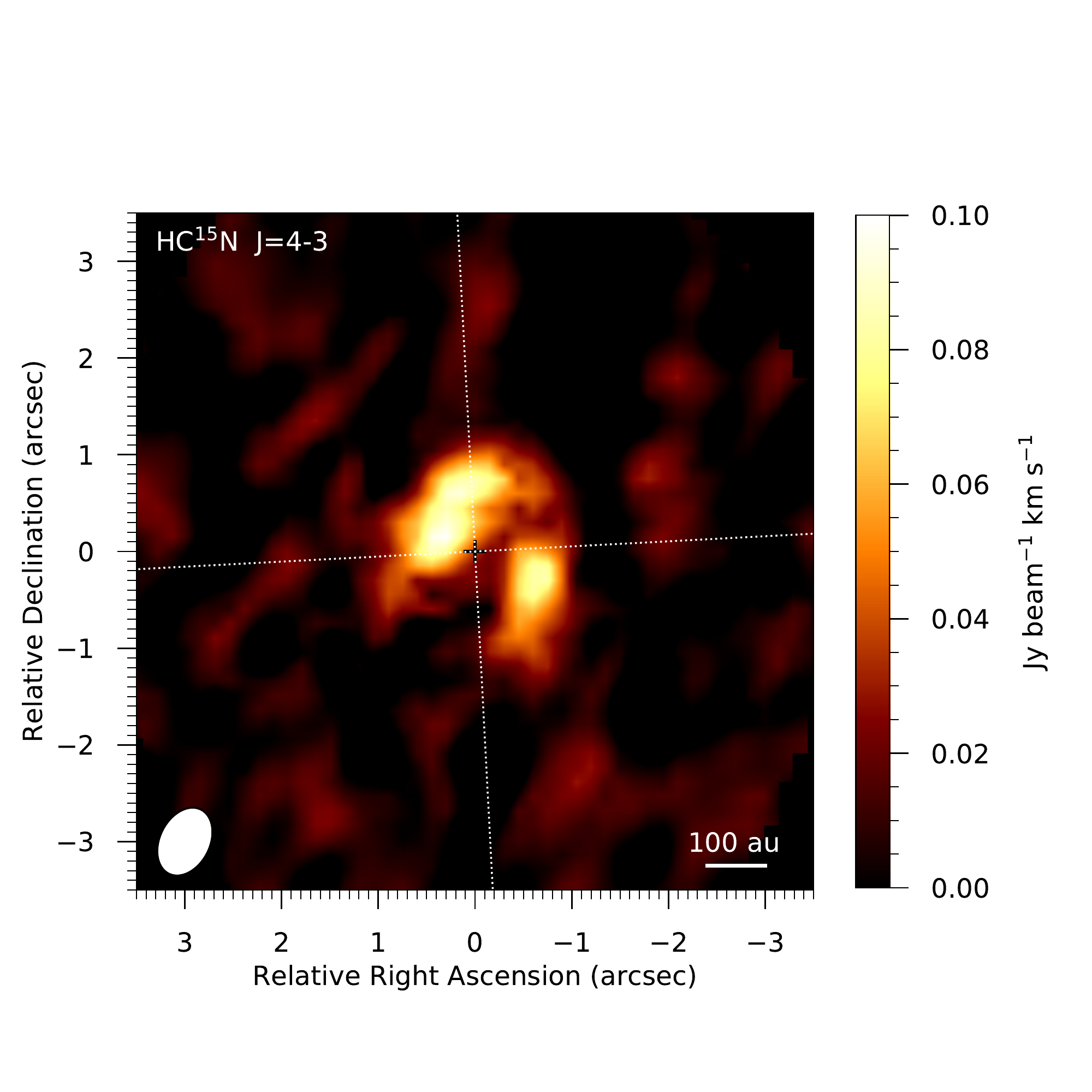}
\includegraphics[trim={0cm 1.5cm 0cm 2.5cm},clip,width=0.5\hsize]{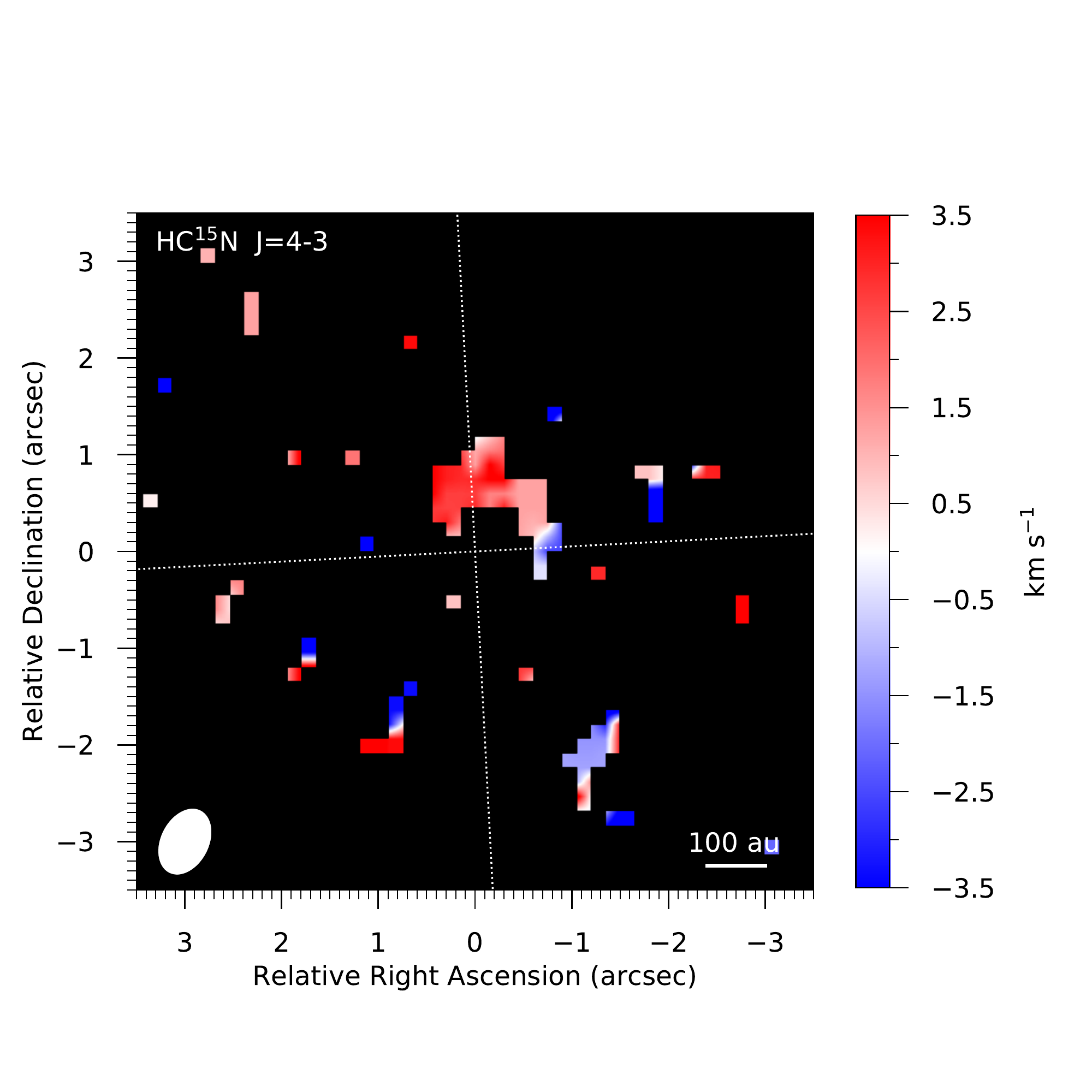}
\caption{The integrated intensity maps (left) and the intensity-weighted velocity (right) for the \ce{HCO+} $J=4-3$ (top), 
\ce{H^{13}CO+} $J=4-3$ (middle) and \ce{HC^{15}N} $J=4-3$ (bottom) transitions 
generated using the Keplerian masks shown in Figures~\ref{figure1}, and A.1.
The S/N of the integrated intensity maps are 20, 3 and 3 respectively. 
The intensity-weighted velocity maps were made using a 3$\sigma$ clip in the channel maps. 
The dotted white lines mark the major and minor axes of the disk.}
\label{figure3}
\end{figure*}


The integrated intensity maps and the intensity-weighted velocity maps for \ce{HCO+}, \ce{H^{13}CO+} and \ce{HC^{15}N} are
presented in Figure~\ref{figure3}. 
The moment 0 maps were made using a Keplerian mask and the
moment 1 maps were made using a 3$\sigma$ clip.
The \ce{HCO+} intensity-weighted velocity map has a clear Keplerian velocity pattern on large spatial scales and 
both the \ce{H^{13}CO+} and \ce{HC^{15}N} intensity-weighted velocity maps are consistent with this.

\begin{figure*}
\centering
\includegraphics[trim={0 0.5cm 0 1cm},clip,width=0.45\hsize]{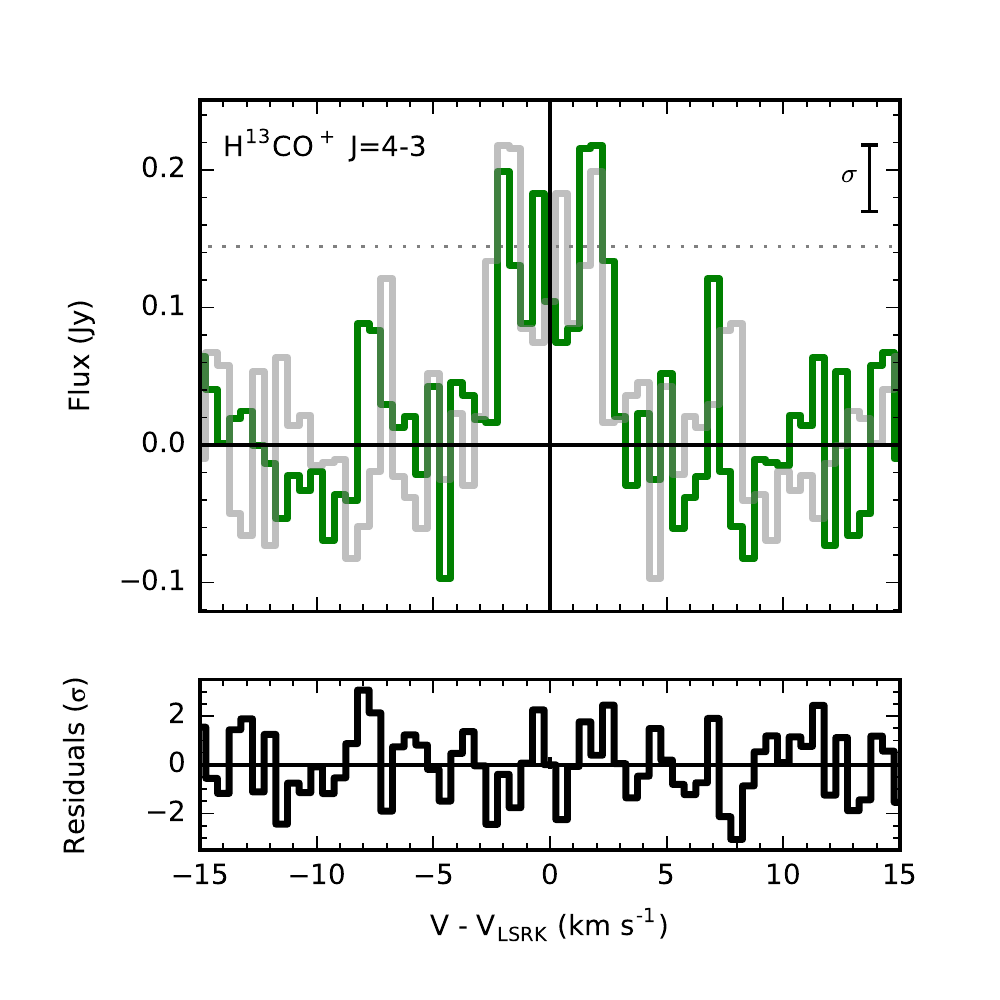}
\includegraphics[trim={0 0.5cm 0 1cm},clip,width=0.45\hsize]{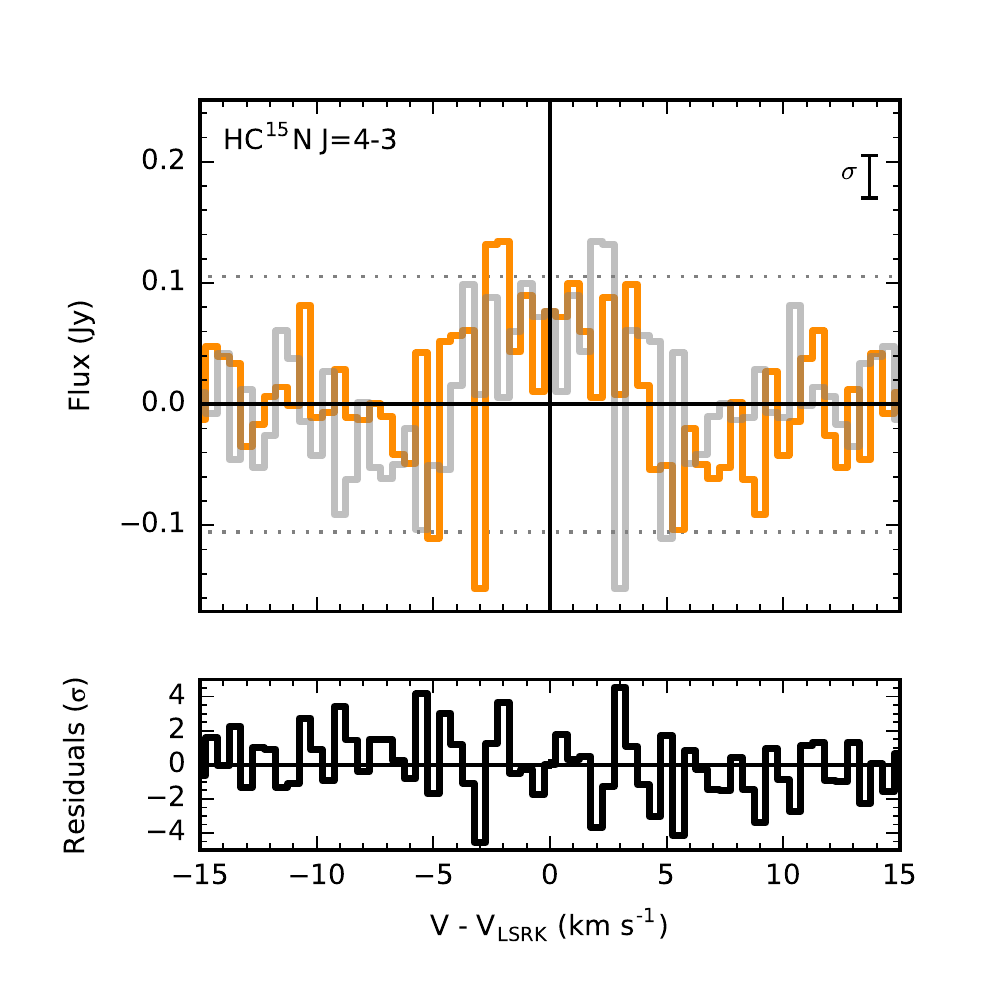}
\caption{Line profiles of the \ce{H^{13}CO+} (left) and \ce{HC^{15}N} (right) 
emission at a velocity resolution of 0.5~km~s$^{-1}$ and mirrored about the source velocity. 
The \ce{H^{13}CO+} line profile is extracted from within the region encompassing the 
$3\sigma$ extent of the \ce{HCO+} integrated intensity and the \ce{HC^{15}N} extraction region 
is an ellipse centred on source and chosen by eye.
The residuals from the mirrored line profiles are also shown.}
\label{figure4}
\end{figure*}

Figure~\ref{figure4} shows the \ce{H^{13}CO+} and \ce{HC^{15}N} line profiles mirrored about the source velocity.
The \ce{H^{13}CO+} line profile was extracted from within the region defined by the 
3$\sigma$ contour of the \ce{HCO+} integrated intensity (see Figure~\ref{figure3}).
The extraction region for the \ce{HC^{15}N} line profile was chosen by eye due to the low S/N.
An ellipse that encompassed the extent of the emission in the integrated intensity map and centred on source was used.   
The line profiles cover a velocity range of $\pm~15$~km~s$^{-1}$ about the source velocity 
and are shown at a velocity resolution of 0.5~km~s$^{-1}$.
They reach a S/N of 5 and 3 with a 1$\sigma$ noise level of 0.048~Jy and 0.035~Jy, respectively.
The residuals of the mirrored line profiles were calculated to show that the emission 
is not significantly asymmetric and to further confirm the Keplerian nature of the emission. 

\subsection{Deprojection and azimuthal averaging}

To better quantify the radial structure of the emission from each 
molecular transition, the Keplerian-masked integrated intensity maps were deprojected and azimuthally averaged.
This is an effective method to increase S/N assuming that the emission is azimuthally symmetric \citep[e.g.,][]{2016ApJ...832..204Y}.
All pixels in each map were placed into radial bins depending on the deprojected radius, and the average value per bin calculated. 
The bin size chosen was 0\farcs3 because this is two times the pixel size of the images and is approximately half of the beam size.    
The recovered fluxes and trends in the resulting radial profiles are consistent with radial profiles generated from un-clipped 
and un-masked integrated intensity maps. 

\begin{figure*}
\centering
\includegraphics[trim={1.5cm 1cm 2.5cm 1cm},clip,width=\hsize]{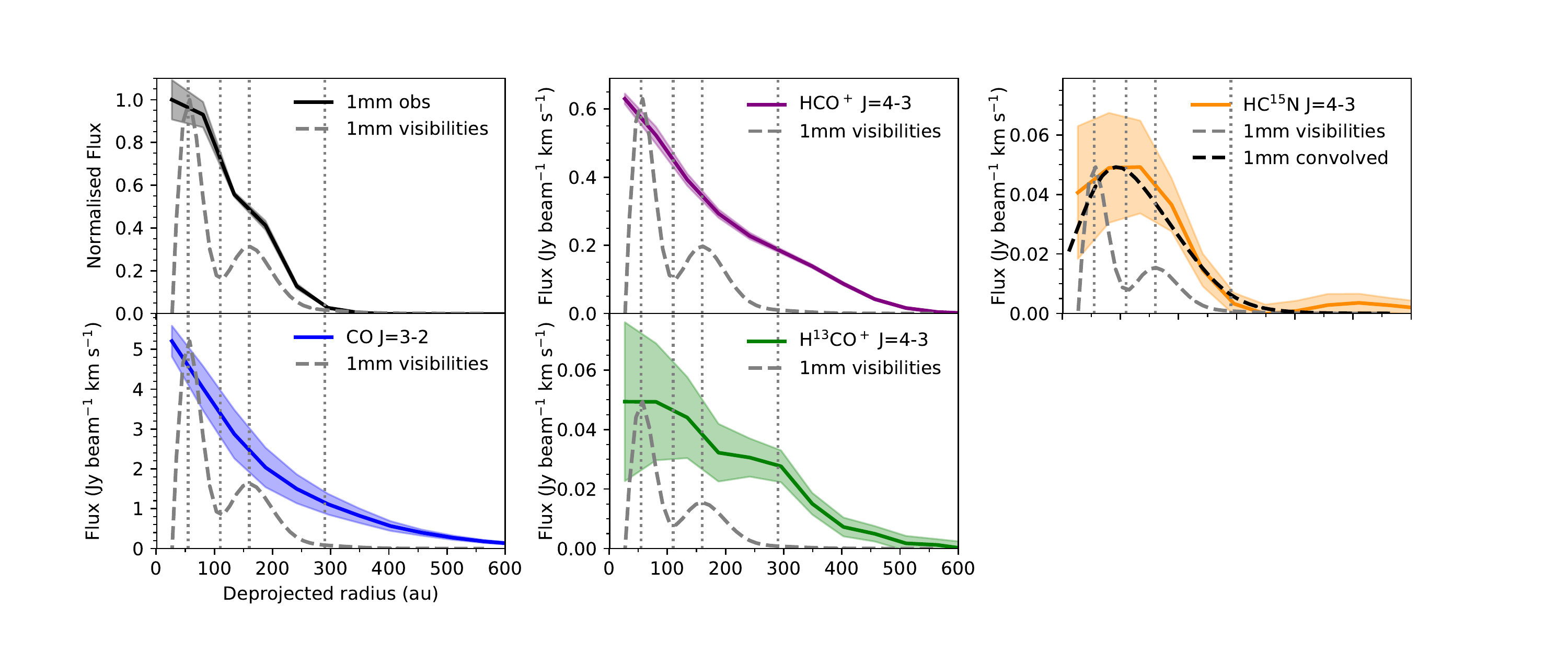}
\caption{Deprojected and azimuthally-averaged radial profiles of the continuum emission, and \ce{CO}, \ce{HCO+}, \ce{H^{13}CO+}, \ce{HC^{15}N}
line emission ratio (from top to bottom, then left to right). 
The coloured shaded region on each profile represents the errors (see text for details).
The vertical dashed grey lines highlight the rings and gaps in the mm-dust continuum profile and the dashed grey profile is the best-fit model 
of the continuum $uv$ data from \citep{2016ApJ...831..200W} and is normalised to the peak value in each plot.
The dotted black line on the \ce{HC^{15}N} plot is the continuum model convolved with a 50~au beam from \citep{2016ApJ...831..200W}
and is normalised to the peak value in each plot.}
\label{figure5}
\end{figure*}

The radial intensity profiles for \ce{CO}, \ce{HCO+}, \ce{H^{13}CO+}, and \ce{HC^{15}N} are shown in Figure~\ref{figure5}.
The associated errors are the standard deviation of the pixel values in each bin divided by the 
square root of the number of beams per annulus \citep[see e.g.][]{2018A&A...614A.106C}.
Also plotted is the mm-dust radial profile from \citet{2016ApJ...831..200W} where the image was generated using super-uniform weighting in CLEAN. 
The continuum profile shown by the grey dashed line in Figure~\ref{figure5} is that derived from modelling the emission in the $uv$ domain. 
The profile has rings which peak at 55 and 160~au with a gap at 110~au and an additional shoulder of emission at 290~au. 
Although these data were lower spatial resolution than that presented in \citet{2017A&A...597A..32V}, modelling the continuum emission 
in the $uv$ domain successfully predicted the presence and location of the outer ring(s) that were subsequently imaged in that work 
\citep[see also][]{2016ApJ...818L..16Z}.

The radial profiles of the \ce{CO} and \ce{HCO+} emission show no 
deviation from a monotonically decreasing radial distribution.  
As discussed previously, the \ce{CO} $J=3-2$ emission suffers from missing emission about the source velocity, from 3.4 to 5.7~km~s$^{-1}$; 
therefore, the radial profile that we derive likely underestimates the CO flux from the disk. 
The profile is centrally peaked because the observations do not have the spatial resolution to resolve the inner CO cavity 
detected in CO ro-vibrational lines \citep{VanderPlas2009}. 
The \ce{HCO+} $J=4-3$ is also expected to be optically thick and the radial 
profile is also centrally peaked.
In comparison, the more optically thin tracers, \ce{H^{13}CO+} and \ce{HC^{15}N}, show a different radial behaviour. 

The \ce{H^{13}CO+} $J=4-3$ radial profile is also centrally peaked but 
the intensity does not fall off as steeply as the profile for \ce{HCO+}. Additionally, the intensity profile appears to have a step-like morphology compared to the
smooth \ce{HCO+} profile.
These changes in slope coincide approximately with the 
locations of the second and third continuum emission rings.  

Comparatively, the \ce{HC^{15}N} profile appears to have a ring-like morphology with the peak in emission occurring at $\approx 100$~au. 
This is consistent with the moment 0 map shown in Figure 2, but due to the low S/N of the data
a centrally peaked smooth radial intensity profile cannot be ruled out. 
Also plotted in Figure 4 with a black dashed line is the mm-dust radial profile from \citet{2016ApJ...831..200W} convolved with a 50~au beam. 
The \ce{HC^{15}N} follows the mm-sized dust distribution at the spatial resolution of the currently available observations.
Further analysis of the \ce{HC^{15}N} emission is not conducted here due to the unavailability of observations of line emission from the 
main isotopologue, HCN.

\begin{figure*}
\centering
\includegraphics[trim={1.5cm 0cm 2.5cm 0cm},clip,width=\hsize]{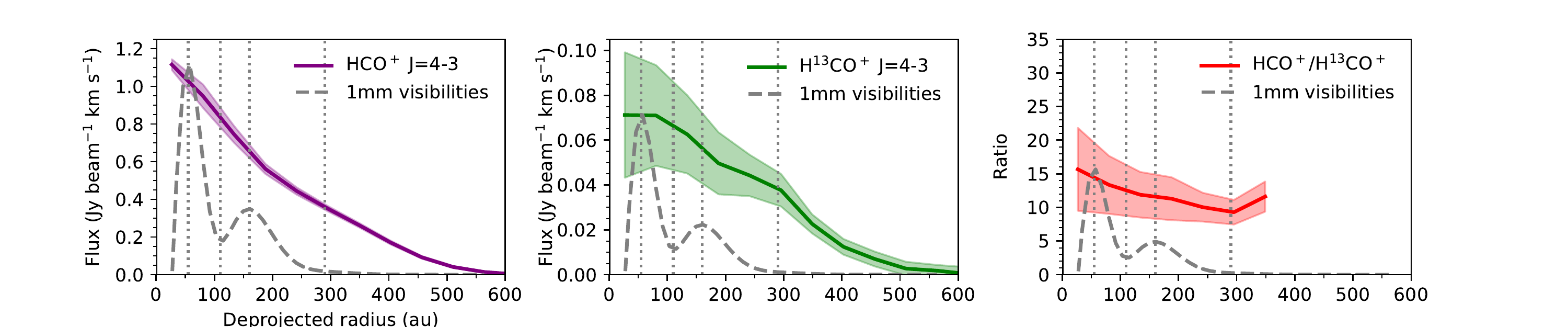}
\caption{Deprojected and azimuthally-averaged radial profiles of the \ce{HCO+} and \ce{H^{13}CO+} line emission smoothed to a common beam
and the subsequent \ce{HCO+}/\ce{H^{13}CO+} ratio (from left to right). 
The coloured shaded region on each profile represents the errors (see text for details).
The vertical dashed grey lines highlight the rings and gaps in the mm-dust continuum profile and the dashed grey profile is the best-fit model 
of the continuum $uv$ data from \citep{2016ApJ...831..200W} and is normalised to the peak value in each plot.}
\label{figure12}
\end{figure*}

The \ce{HCO+} to \ce{H^{13}CO+} integrated intensity ratio across the disk 
is shown in Figure~\ref{figure12}.
To directly compare the \ce{HCO+} and \ce{H^{13}CO+} emission from the disk, the 
channel maps have been smoothed to the same beam size using the CASA task, \texttt{imsmooth}. 
The new beam is 0\farcs7~$\times$~0\farcs6 with a position angle of -26\degree.
We also tried the same analysis using a \textit{uvtaper} when CLEANing the lines
to enforce a common beam and found that the results were consistent with using \texttt{imsmooth}.
New moment maps were generated and radial profiles from the resulting 
integrated intensity maps used to calculate the \ce{HCO+} to \ce{H^{13}CO+} integrated intensity ratio across the disk. 
The smoothed radial profiles are also shown in Figure~\ref{figure12}. 
The plotted ratio is truncated at the point where the \ce{H^{13}CO+} profile falls below the noise level.  
The computed errors have been propagated from the standard error for each profile generated during azimuthal averaging.
The expected ratio is $\approx$~69 for the case where both lines are optically thin, 
reflecting the underlying elemental ratio for \ce{^{12}C}/\ce{^{13}C} in the ISM \citep[e.g.,][]{1999RPPh...62..143W}.  
The observed ratio decreases slightly with radius but is constant across the disk within the error bars, and has an average value of $12~\pm~3$.
This is much lower than the canonical value and confirms that at least the \ce{HCO+} emission is optically thick.

\section{Chemical Modelling}
\label{chemicalmodelling}

In this section we investigate the origin of the radial distribution of the 
detected \ce{H^{13}CO+} emission and the \ce{HCO+}/\ce{H^{13}CO+} 
integrated intensity ratio. We model the line emission 
from both molecules to explore two hypotheses i) whether or not 
chemistry alone can explain the radial emission profiles, 
or, ii) if modifications to the underlying disk structure need to be invoked (e.g., gas cavities).
We note that we do not attempt to model the \ce{HC^{15}N} emission since we 
do not have complementary observations of the main isotopologue \ce{HCN}. 
However, we discuss this detection and related chemistry in Section 6.

For the underlying physical disk structure we use the publicly available  
HD~97048 disk structure from the DIscANAlysis project 
(DIANA)\footnote{The full DIANA models are publicly available: \url{http://www-star.st-and.ac.uk/~pw31/DIANA/DIANAstandard/}}.  
The aim of this large collaboration is to generate disk models that can
reproduce multi-wavelength (from UV to sub-mm) observations of sources for which data are available 
(see \citealt{2016A&A...586A.103W}, \citealt{2017A&A...607A..41K} and \citealt{2019PASP..131f4301W}, for further details on the models). 
The underlying model assumes a parametric disk model, and for which the gas temperature is self-consistently 
calculated considering the balance between cooling and heating mechanisms for the gas. 
The resulting model fits the SED of the source and aims to reproduce the integrated flux from far-IR gas emission lines. 
The underlying disk structure is assumed to be smooth; however, to better fit SEDs at near to mid-IR wavelengths, models 
are adapted to include a dust (and gas) cavity if required.  
Generally, models have not been generated to account for dust gaps and rings as seen in spatially-resolved mm dust continuum images.

\begin{figure*}
\centering
\includegraphics[trim={0.0cm 0.0cm 0.0cm 0.0cm},clip,width=.9\hsize]{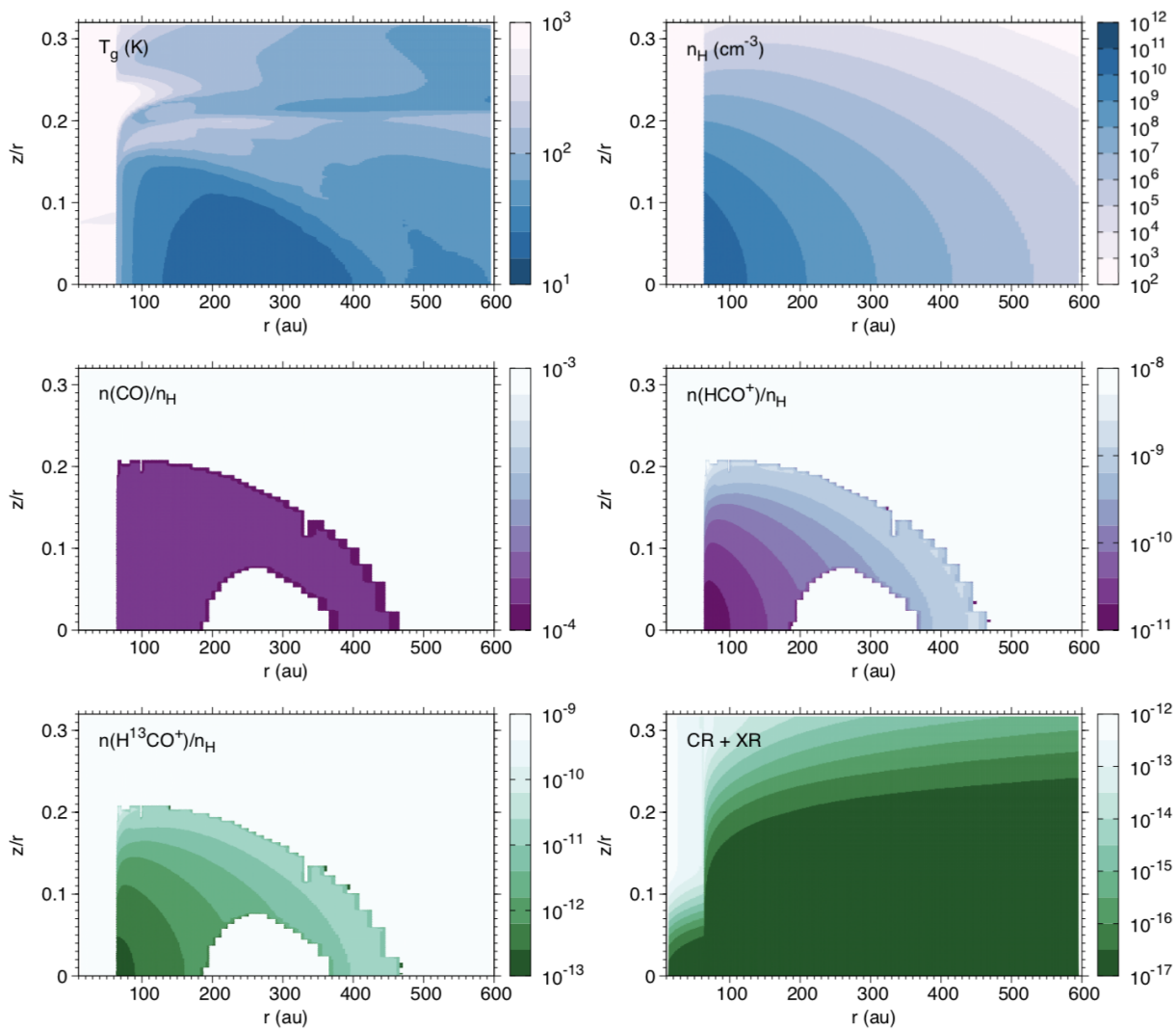}
\caption{The physical disk structure of the HD~97048 protoplanetary disk 
and the results from the analytical chemical model. 
Quantities shown are the gas temperature (top left), total H nuclei number density (top right), 
CO number density relative to H nuclei (middle left), 
\ce{HCO+} number density relative to H nuclei (middle right), 
\ce{H^{13}CO+} number density relative to H nuclei 
(bottom left) and the sum of the cosmic ray and X-ray ionisation rates (bottom right).}
\label{figure6}
\end{figure*}

We use the gas density, gas temperature, and ionisation structure from the DIANA model of HD~97048.
This model has a total disk (gas plus dust) mass of $\approx~0.1~{M_\odot}$.  
Assuming a global gas-to-dust mass ratio of $\sim 100$, this disk mass is 
consistent with the upper limit for the dust mass derived from the (sub)mm continuum emission 
observed with ALMA and corrected for the revised distance to the source \citep[$\lesssim 0.09 \mathrm{M_{J}}$;][]{2016ApJ...831..200W}.
The ionisation rate throughout the disk is the sum of both the cosmic-ray and X-ray ionisation rates.
Maps of the disk physical structure as a function of disk radius and height
are shown in Figure~\ref{figure6}. 
The best-fit DIANA model has a cavity out to 65~au in both the gas and the dust and 
a small inner ring of gas and dust out to 6~au. 
We note here that the size of this cavity is greater than that observed with ALMA in the 
mm-dust continuum ($\approx$50~au).


Due to chemistry, the abundance distribution of \ce{HCO+} will not have a constant value relative 
to the underlying gas density as is commonly assumed in parametric models. 
Following \citet{2015ApJ...807..120A}, we analytically derive the \ce{HCO+} abundance, 
$n(\ce{HCO+})$~cm$^{-3}$, throughout the disk under a few acceptable and tested assumptions. 
Because the chemical timescale in a disk is significantly shorter than the 
disk lifetime we assume steady state, i.e., $$ \frac{dn(\ce{HCO+})}{dt}=~R_{f}~-~R_{d}~= 0,$$
where $R_{f}$ and $R_{d}$ are the rates of \ce{HCO+} formation and destruction, respectively. 
The main formation reaction for \ce{HCO+} is via \ce{CO},
\begin{align}
\ce{H3+} + \ce{CO} &\longrightarrow \ce{HCO+} + \ce{H2}, \nonumber 
\end{align}
and the primary destruction reaction is recombination with an electron,
\begin{align} 
\ce{HCO+} + \ce{e-} &\longrightarrow \ce{H} + \ce{CO} \nonumber . 
\end{align}
We ignore the destruction of \ce{HCO+} via grain-surface recombination as this is only dominant in the 
inner $<$ 10~au of the disk in the dense midplane where dust grains become the main charge carrier and 
we do not model this region of the disk. 
We also assume that \ce{HCO+} is the dominant cation in the molecular region of the disk as has been confirmed by full chemical 
models \citep[e.g.,][]{2012ApJ...747..114W, 2015ApJ...807..120A}. 
The resultant abundance of \ce{HCO+} is given by;
$$ n(\ce{HCO+}) = \frac{k_3 ~n(\ce{CO}) ~ n(\ce{H_3^+})}{k_4 ~ n(\ce{e-})}.$$
The reaction rate coefficients, $k_i$, are given in Table~\ref{table3}, 
where $i$ is the number of the reaction as compiled by \citet{2015ApJ...807..120A}.

\begin{table*}
\centering
\caption{Reaction rate coefficients from UMIST {\sc Rate}12$^{1}$ used in the analytical formula from \citet{2015ApJ...807..120A}. \label{table3}}
\begin{tabular}{clcc}
\hline \hline
$i$ & {Reactions}                                                        & $\alpha$ & $\beta$  \\ \hline
1   & $\ce{H2} + \mathrm{crp} \longrightarrow \ce{H2+} + \ce{e- }$       & -        & -        \\
2   & $\ce{H2+} + \ce{H2} \longrightarrow \ce{H3+} + \ce{H}  $           & 2.08(-9) & 0.0        \\
3   & $\ce{H3+} + \ce{CO} \longrightarrow \ce{HCO+} + \ce{H2}$           & 1.61(-9) & 0.0      \\
4   & $\ce{HCO+} + \ce{e} \longrightarrow \ce{H} + \ce{CO}  $            & 2.40(-7) & -0.69    \\
5   & $\ce{H3+} + \ce{e} \longrightarrow \ce{H2} + \ce{H}$~or~H + H + H  & 7.20(-8) & -0.5     \\ \hline
\end{tabular}
\begin{tablenotes}
\footnotesize
\item{Rate coefficients are given in the form of $k=\alpha\times(T/300.0)^{\beta}~\mathrm{cm}^{3}\mathrm{s}^{-1}$ and
$A(B)$ stands for $A~\times~10^{B}$}
\item{$^1$ \url{http://udfa.ajmarkwick.net/index.php}, \citep{2013A&A...550A..36M}}
\end{tablenotes}
\end{table*}

We parameterise the \ce{CO} abundance distribution as described by \citet{2014ApJ...788...59W}.  
\ce{CO} is present in the disk with a relative abundance of $n$(CO)/$n_{\ce{H}}=1.4\times10^{-4}$
if the temperature is greater than 20~K (the approximate freezeout temperature of \ce{CO}) and if the 
vertical column density of \ce{H2} (integrated vertically downwards from the disk surface) 
is greater than $1.3~\times~10^{21}$~cm$^{-2}$. 
Here, $n_{\ce{H}}$ is the total number density of hydrogen nuclei.
This simple prescription takes into account the two primary destruction pathways for CO: freezeout onto dust grain surfaces 
in the midplane, and photodissociation in the atmosphere. 
The column density upper limit comes from modelling work by \citet{2009A&A...503..323V} and observations of HD~163296 by \citet{2011ApJ...740...84Q}.
The \ce{H_3^+} and electron abundances are given by
$$ n(\ce{H_3^+}) = \frac{1}{2} ~ n_{\ce{H}} ~ \frac{\zeta}{k_5~n(\ce{e-})~+~k_3~n(\ce{CO})},$$ and
$$ n(\ce{e-}) = \sqrt{\frac{\zeta~n_{\ce{H}}}{2~k_4}}, $$ 
\citep{2015ApJ...807..120A}.
Here $\zeta$ is the ionisation rate from the DIANA model. 
These equations are determined by solving the rate equations for each species under the steady state assumption. 

The \ce{H^{13}CO+} abundance is determined by simply assuming that 
\ce{H^{12}CO+}/\ce{H^{13}CO+}=\ce{^{12}C}/\ce{^{13}C}~=~69 \citep[e.g.,][]{1999RPPh...62..143W}.
Although isotope fractionation processes (e.g., fractionation reactions and isotope-selective photodissociation) will 
cause the \ce{^{12}CO}/\ce{^{13}CO} and \ce{H^{12}CO+}/\ce{H^{13}CO+} abundance ratios to vary throughout 
the disk \citep{2009ApJ...693.1360W}, developing and running a full chemical model that includes fractionation
is beyond the scope of this work and best addressed in a future dedicated study. 
In a previous study, \citet{2009ApJ...693.1360W} show that the $n$(\ce{H^{12}CO+})/$n$(\ce{H^{13}CO+}) ratio has a value of $\approx 40-60$ in the layer 
in the inner disk ($\lesssim 30$~au) at which \ce{HCO+} reaches the highest fractional abundance ($\sim 10^{-9}$).
The resulting abundance distributions as a function of disk radius and height  
for \ce{CO}, \ce{HCO+} and \ce{H^{13}CO+} are shown in Figure~\ref{figure6}.

Synthetic channel maps were generated using LIME version 1.6 \citep[LIne Modelling Engine;][]{2010A&A...523A..25B}
using the molecular data files for \ce{HCO+} and \ce{H^{13}CO^+} from the LAMDA molecular 
database\footnote{LAMDA data files are available online at: \url{http://home.strw.leidenuniv.nl/~moldata/}}. 
To check whether or not the assumption of LTE was appropriate, we calculated the critical density of the 
\ce{HCO+} $J=4-3$ transition using the collisional rates from \citet{1999MNRAS.305..651F}.
The critical density was determined to be $10^{6}$~cm$^{-3}$ in the temperature range from to 30 to 100~K.
In the region of the disk model where the \ce{HCO+} is present the gas density is above 
this value (see Figure~\ref{figure6}); therefore, it was confirmed that LTE is an appropriate assumption. 
The models were made using the same position and inclination angles of the source.
The resulting image cubes were then smoothed to the same Gaussian beam as the \ce{HCO+} and \ce{H^{13}CO+} observations.
The simulated channel maps were used to generate azimuthally-averaged radial profiles
from the resulting integrated intensity maps and then directly compared with the observations.
For the \ce{H^{12}CO+}/\ce{H^{13}CO+} integrated intensity ratio the images were smoothed to 
the same common beam as described for the observational dataset. 
We run a range of models step-by-step altering the initial model to find a best by-eye fit to the 
observed radial profiles. The specifics of the models are summarised in Table 3.
The resulting modelled radial intensity profiles for \ce{HCO+} and \ce{H^{13}CO+}, and the line ratio,
are shown in Figure~\ref{figure7} where the coloured lines are the observations and the black lines are the models.

\subsection{Fiducial Model Results}

Model 1 is the fiducial model as described above and this model is found to overestimate the magnitude 
of the emission for both the \ce{HCO+} and \ce{H^{13}CO+} lines. 
The difference between the modelled and observed peak flux for \ce{HCO+} and \ce{H^{13}CO^+} 
are around a factor of 1.3 and 2.4, respectively. 
The morphology of the modelled emission also does not match the observed profiles.  
The latter are centrally peaked whereas the modelled profiles peak in emission at $\approx 100$~au.
The resulting line ratio also does not 
exhibit the same radial behaviour as the observation as it increases radially from 6 to 26.
 
\subsection{Modifications to the Fiducial Model}

Since the fiducial model is a poor fit to the observations we now modify this model 
step by step in order to better reproduce the data. 
We first aim to fit the radial profiles beyond a radius of $\approx$100~au. 
In Model 2 the abundance of \ce{HCO+} was multiplied by a factor, 
$\kappa_{\ce{HCO^{+}}}$, of 0.2 relative to the abundance calculated using the analytical chemical model. 
This factor was determined by gradually reducing the multiplication factor in steps of 0.1 in each test, i.e., [0.9, 0.8, ...], until an
acceptable by-eye fit to the \ce{HCO+} was found.
The multiplicative factor $\kappa_{\ce{H^{13}CO^{+}}}$ is additionally applied to the \ce{H^{13}CO+} which alters the
 $n$(\ce{H^{12}CO+})/$n$(\ce{H^{13}CO+}) ratio in the disk. For Model 2 $\kappa_{\ce{H^{13}CO^{+}}}$ is 1 i.e. 
$n$(\ce{H^{12}CO+})/$n$(\ce{H^{13}CO+}) is 69.
A comparison with the observed profiles shows that good agreement with the radial integrated 
intensity profile for \ce{HCO+} emission is now obtained beyond 100~au.
The overall decrease in abundance of a factor of 5 that is required here indicates that either the gas surface 
density is overestimated by this factor in the disk, or that the disk temperature of the emitting layer is over-estimated, 
or that the analytical chemical model describing \ce{HCO+} is over-predicting the fractional abundance.
However, in this model the \ce{H^{13}CO+} peak flux is now under-predicted. 
The resulting \ce{H^{12}CO+}/\ce{H^{13}CO+} integrated intensity ratio still does not exhibit the same radial 
behaviour as the observations and is increasing radially from 16 to 44.

\begin{figure*}
\includegraphics[trim={.0cm 0.0cm 0.0cm 0.0cm},clip,width=\hsize]{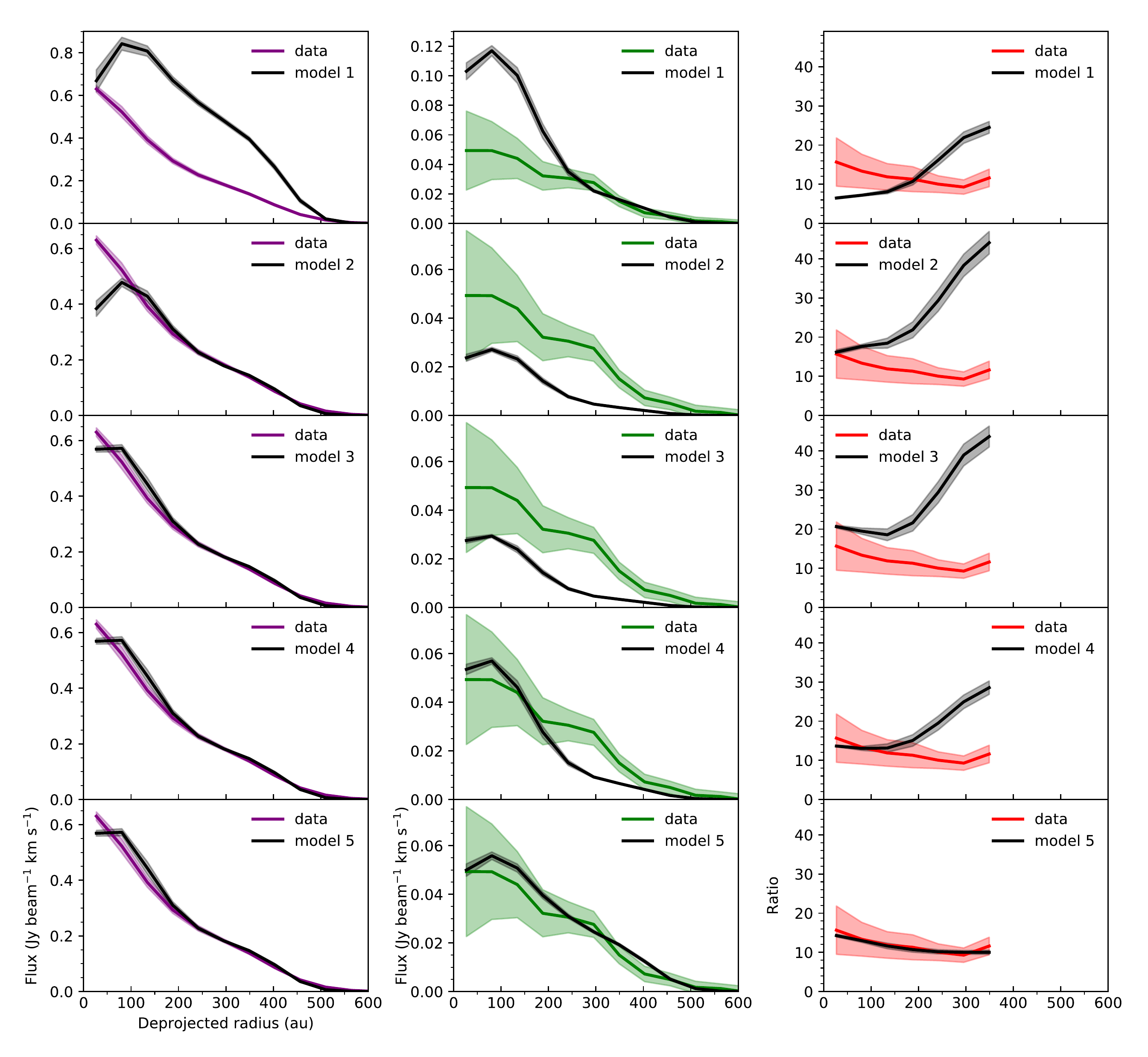}
\caption{Deprojected and azimuthally-averaged radial integrated intensity profiles for \ce{HCO+}, \ce{H^{13}CO+} and 
the \ce{HCO+}~/\ce{H^{13}CO+} ratio from the observations (in colour) and the LIME models (black). 
The coloured shaded region on each profile represents the errors (see text for details).
The specifics of each model are detailed in Table 3 and Section 5.}
\label{figure7}
\end{figure*}

Next we try to reconcile the \ce{HCO+} model and observations in the inner disk and apply the same changes to the \ce{H^{13}CO+} in order to retain their 
abundance ratio throughout. 
In Model 3 we modify the disk structure by hand to include gas in the cavity from $\approx6$ to $\approx65$~au.
The gas density, temperature, and molecular abundances within the cavity are 
set equal to the values at the inner edge of the model gas disk. 
An LTE model with gas in the cavity and with 0.2 times the \ce{HCO+} abundance ($\eta_{cavity}$) used in Model 2 
reproduces well the observed radial profile (see Figure~\ref{figure7}). 
The continued disagreement with the \ce{H^{13}CO+} observations and models shows that the 
initial assumption of a globally constant value for $n$(\ce{H^{12}CO+})/$n$(\ce{H^{13}CO+}) of 69 may not be appropriate.

For the \ce{H^{13}CO+}, further modifications to this model are required to better reproduce the observations. 
We alter the disk $n$(\ce{H^{12}CO+})/$n$(\ce{H^{13}CO+}) abundance ratio globally by increasing  
$n$(\ce{H^{13}CO+}) relative to the values in Model 3. 
The best by-eye fit for $\kappa_{\ce{H^{13}CO^{+}}}$ was found to be 2 relative to the abundances in Model 2. 
This results in a 
global disk $n$(\ce{H^{12}CO+})/$n$(\ce{H^{13}CO+}) abundance ratio of 35. This is not unrealistic as chemical models that include
chemical isotope fractionation processes do predict a decrease in $n$(\ce{^{12}CO})/$n$(\ce{^{13}CO}) and $n$(\ce{H^{12}CO+})/$n$(\ce{H^{13}CO+}) by at most a 
factor of two in the midplane \citep{2009ApJ...693.1360W,2014A&A...572A..96M}. 
This model provides a much improved fit in the inner disk within 200~au in the \ce{H^{13}CO+}, but there is still not enough \ce{H^{13}CO+} emission
in the outer disk.

In Model 5 we decrease the $n$(\ce{H^{12}CO+})/$n$(\ce{H^{13}CO+}) ratio in the outer disk ($>$200~au) further to try and fit the 
\ce{H^{13}CO+} observations. 
The best by-eye fit for $\kappa_{\ce{H^{13}CO^{+}}}$ was found to be 6 relative to the abundances in Model 2 resulting in a 
$n$(\ce{H^{12}CO+})/$n$(\ce{H^{13}CO+}) ratio of 14 in the outer disk . 
This suggests the presence of two carbon chemical fractionation regimes, traced in \ce{H^{13}CO^+} in the HD~97048 disk. 
The feasibility of this scenario and possible explanations will be discussed in Section 6.3.

\begin{table*}
\centering
\caption{The different models as described in Section 5 along with the parameters that are varied in each model.
The ticks and crosses denote if the model fits the specified observation.}
\begin{tabular}{ccccccccc}
\hline\hline
Model & $\kappa_{\ce{HCO^{+}}}$  & $\kappa_{\ce{H^{13}CO^{+}}}$                     & Gas in cavity & $\eta_{cavity}$            & \underline{$n$(\ce{HCO+})}   & \ce{HCO+}  & \ce{H^{13}CO^{+}} &\underline{\ce{HCO+}}  \\
 &   &                     &                                                        &                   & 	$n$(\ce{H^{13}CO^+})		&			&				& \ce{H^{13}CO^+}	\\ \hline

1     & -                             & -                 				 				& no            	& -                      	& 69   						& x &x&x\\
2     & [0.9, 0.8,...,\textbf{0.2}]   & 1			     				 			& no            	& -                      	& 69                    	&x&x& x\\
3     & 0.2                       & 1	                				 		& yes           	& [0.9, 0.8,...,\textbf{0.2}] & 69                        &\checkmark&x&x\\
4     & 0.2                       & 2    				 	    		& yes           	& 0.2                    	& 34.5    					&\checkmark&x&x\\
\multirow{2}{*}{5}                & 0.2 (< 200~au)            & 2 (< 200~au)    					 			& \multirow{2}{*}{yes}       		& \multirow{2}{*}{0.2}                  		& 35 (< 200~au)  				&\multirow{2}{*}{\checkmark}&\multirow{2}{*}{\checkmark}&\multirow{2}{*}{\checkmark}\\                                                         
      & 0.2 (> 200~au)            & [2, 3,...,\textbf{6}](> 200~au)   				& 					&                      		& 11.5 (> 200~au)               &&&\\               \hline                                          
\end{tabular}
\end{table*}

We also show the residual channel maps (observations minus Model 5) 
in Appendix C for \ce{HCO+} and \ce{H^{13}CO+}, respectively.
From the channel maps we are primarily under-predicting (purple contours) the flux from the \ce{HCO+} line 
despite the reasonably good fit to the integrated intensity profile.
These residuals may be due to the model emission originating from a lower height in the disk than the observations. 
This would mean that the \ce{HCO+} gas is optically thick higher in the disk than our model predicts
such that the model is either underestimating the disk temperature and hence the disk scale height (i.e., it is more flared than predicted), 
or that the disk is more massive than suggested by the DIANA model.
The \ce{H^{13}CO+} residuals are insignificant.

We also checked the impact on the modelled line emission when including the effect of dust opacity.
The aim was to check whether or not this arises in a suppression of emission in the inner region. 
We assumed a gas-to-dust mass ratio of 100 and used the dust opacity table from 
\citet{1994A&A...291..943O} assuming $10^6$ years of coagulation and thick ice mantles.
This is reasonable for the age of this disk \citep[$\approx~3$~Myr,][]{2006Sci...314..621L}.
The continuum was subtracted from the model images using the CASA function \texttt{imcontsub}.
The resulting radial profiles differed by 10 to 20\% only from the models where we neglect dust opacity
and thus are within the errors of the data.
Hence, the emission in the inner disk is not significantly suppressed due to dust 
opacity. The explanation for this is likely because we are not resolving molecular emission 
from the innermost region where dust opacity has the maximal impact on the emergent line emission  
\citep[e.g.,][]{2015ApJ...810..112O,2016PhRvL.117y1101I}.

To investigate the effect of the gas temperature on the \ce{HCO+} and \ce{H^{13}CO+}
line emission we ran a set of exploratory models.
We found that reducing the gas temperature by a factor of
2 throughout the disk led to a good by-eye fit to the observations in the inner disk 
($<$ 200 au) using the canonical abundance ratio.  However, a significant 
increase in the \ce{H^{13}CO+} abundance relative to \ce{HCO+} in the outer disk is still 
required to better reproduce the emission here.  This supports our claim of the 
presence of two carbon fractionation regimes in this protoplanetary disk, and also 
supports enhanced fractionation in the outer disk due to the disk being colder than 
previously inferred from previous modelling work.  A more robust determination of the gas 
temperature structure of this source will require high spatial resolution observations of 
multiple lines of molecular tracers of gas temperature.

\section{Discussion}
\label{discussion}

We have presented the first detections of \ce{H^{13}CO+} and \ce{HC^{15}N} in the 
HD~97048 protoplanetary disk.
We compare the radial emission profiles of these 
tracers with the mm continuum emission profile and the line emission from \ce{CO} and \ce{HCO+}.
We find that the optically thin tracers, \ce{H^{13}CO+} and \ce{HC^{15}N}, 
appear to exhibit radial variations in their integrated intensity profiles in contrast to the optically thick tracers, \ce{CO} and \ce{HCO+}.
Contrary to the \ce{H^{13}CO+} $J=4-3$ data, the \ce{H^{13}CO+} $J=4-3$ radial intensity profile shows a step-like radial structure. 
The \ce{H^{12}CO+}/\ce{H^{13}CO+} intensity ratio across the disk is constant with an average value of $12\pm3$.
This value is consistent with observations of other disks where 
the disk-integrated \ce{H^{12}CO+}/\ce{H^{13}CO+} ($J=3-2$) flux ratios range 
from 13 to 25 \citep{2010ApJ...720..480O, 2011ApJ...734...98O,2017ApJ...835..231H}.
This low ratio ($<$~69) is consistent with optically thick \ce{HCO+} emission. 
However, consideration of optical depth effects alone cannot explain the data. 
We have shown that radial variations in the abundance ratio of 
\ce{H^{12}CO+} to \ce{H^{13}CO+} are necessary to explain the radial behaviour in the observed intensity ratio.
We find one model that provides good by-eye fit to the data. 
In Model 5 we require an enhancement relative to the canonical ratio of \ce{H^{13}CO+} to \ce{H^{12}CO+} by a factor of 2 in the inner disk (< 200~au) and 
5 in the outer disk (> 200~au).

The \ce{HC^{15}N} intensity profile appears to peak in a ring 
which appears to match well with the mm-sized dust distribution at the same spatial resolution. 
Higher spatial resolution observations are required to confirm this association.  
Without complementary observations of the main isotopologue, HCN, it 
is difficult to investigate further the origin of the \ce{HC^{15}N} emission. 
However, we can rule out suppression of line emission by the dust opacity in the inner disk, within the first dust ring, as the potential origin of a 
molecular ring. 

In the remainder of this Section, 
we explore if isotope-selective chemistry can explain 
the structure seen in the \ce{H^{13}CO+} emission profile and \ce{H^{12}CO+}/\ce{H^13CO+} ratio across the disk.
We also make an estimate on the level of gas depletion in the inner cavity of the disk, 
compare our observations with those towards other 
Herbig Ae/Be disks, and speculate on alternative 
physical/chemical processes that could explain our results.

\subsection{Can isotope-selective chemistry explain the \ce{H^{13}CO+} emission?}

Our best fit model requires multiple \ce{H^{12}CO+}/\ce{H^{13}CO+} fractionation regimes. 
Isotope selective chemistry
could be responsible for the radial intensity profile of the \ce{H^{13}CO+} emission and the observed \ce{H^{12}CO+}/\ce{H^{13}CO+} ratio
but chemical models with isotope selective chemistry only predict a change in the 
\ce{H^{12}CO+}/\ce{H^{13}CO+} ratio over very narrow radial and vertical regions of the disk.  
While detailed chemical modelling of this is beyond the scope of this work, we explore
how isotope selective chemistry could influence the emergent \ce{HCO+} and \ce{H^{13}CO+} line emission.

Isotope selective photodissociation enhances the ratio of \ce{^{12}CO}/\ce{^{13}CO} in the disk atmosphere due to 
the different self-shielding column densities of the different isotopologues \citep{2009A&A...503..323V, 2014A&A...572A..96M}.
The relative abundances of \ce{^{12}CO} and \ce{^{13}CO} are the least affected of the different \ce{CO} 
isotopologues since they are the most abundant. 
However, the shielding column for \ce{^{13}CO} will be reached slightly deeper into the disk atmosphere than that for 
\ce{^{12}CO} \citep{2014A&A...572A..96M}.
Hence, there will exist a layer in the disk surface where \ce{^{12}CO}/\ce{^{13}CO} $> 69$ such that 
\ce{H^{12}CO+}/\ce{H^{13}CO+} is also $> 69$ due to the coupled chemistry between the two species.
In the model of a T Tauri disk in \citet{2014A&A...572A..96M} that includes comprehensive 
isotope-selective chemistry, the peak abundance ratio for \ce{^{12}CO}/\ce{^{13}CO} is only $\approx~2~\times$ the underlying elemental ratio
and only in a narrow layer of the disk atmosphere.
The models of carbon isotope fractionation in the inner 30~au of a T Tauri disk by 
\citet{2009ApJ...693.1360W} show that there is an increase in the isotopologue ratio in \ce{HCO+} relative to 
that for \ce{CO} in the more tenuous and irradiated disk atmosphere 
(i.e., $n(\ce{H^{12}CO+})/n(\ce{H^{13}CO+}$) > $n(\ce{^{12}CO})/n(\ce{^{13}CO})$ with a maximum 
$n(\ce{H^{12}CO+})/n(\ce{H^{13}CO+}$) of $\approx$~110). 
However, this cannot explain the results for HD~97048 because isotope selective photodissociation leads to 
an enhancement of the main isotopologue relative to the rare isotopologue, which acts in the opposite direction to that 
required to explain our observations.

Additionally, there are isotope exchange reactions which enhance the abundance of \ce{^{13}CO} relative to \ce{^{12}CO} and \ce{H^{13}CO+}
relative to \ce{H^{12}CO+} at low temperatures, <35~K and <9~K respectively \citep[as measured by][]{1980ApJ...242..424S}.
Models of the carbon isotope fractionation in protoplanetary disks show that the chemical 
\ce{HCO+} fractionation follows the \ce{CO} fractionation for warm temperatures >~60~K but for lower
temperatures the fractionation is more extreme for \ce{HCO^+} than \ce{CO}.
In the \citet{2009ApJ...693.1360W} models the \ce{^{13}CO}/\ce{^{12}CO} reaches $\approx$~40 in the disk midplane whereas the \ce{H^{13}CO+}/\ce{H^{12}CO+}
ratio is lower $\approx$~30.  
Since the density is the disk is relatively high this process is roughly in chemical equilibrium and can be calculated as follows:

$$\frac{n(\ce{H^{12}CO^+})}{n(\ce{H^{13}CO^+})}~=~\exp{\Big(\dfrac{-9\mathrm{K}}{\mathrm{T_{gas}}}\Big)}~\frac{n(\ce{^{12}CO})}{n(\ce{^{13}CO})}.$$

Figure 8 shows the $n(\ce{H^{12}CO^+})$/$n(\ce{H^{13}CO^+})$ ratio as a function of gas temperature and underling 
$n(\ce{^{12}CO})$/$n(\ce{^{13}CO})$ ratio. The white contours mark the three $n(\ce{H^{12}CO^+})$/$n(\ce{H^{13}CO^+})$ ratios
used in Models 1 to 3, 4 and 5 respectively. The dashed black line marks the lowest gas temperature in the disk model ($\approx$~16~K). 
This calculation shows us that the fractionation required in Model 4 is reasonable, but 
in order to reach the level of fractionation in Model 5, depending on the underlying $n(\ce{^{12}CO})$/$n(\ce{^{13}CO})$ ratio, 
the gas temperature needs to be between 5 and 10~K and this is less than the minimum gas temperature in the disk model.
For chemical fractionation to explain our results the gas temperature in the outer disk needs to be significantly cooler than 
current models predict.

\begin{figure}
\includegraphics[trim={0 0 0 0},clip,width=\hsize]{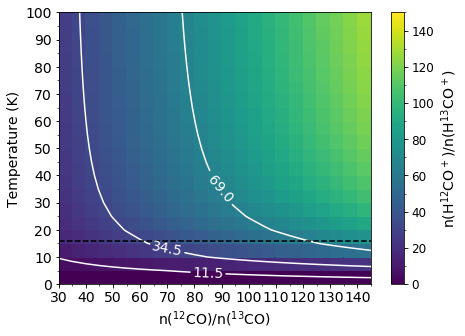}
\caption{The $n(\ce{H^{12}CO^+})$/$n(\ce{H^{13}CO^+})$ ratio as a function of gas temperature and underling 
$n(\ce{^{12}CO})$/$n(\ce{^{13}CO})$ ratio. The white contours mark the three $n(\ce{H^{12}CO^+})$~/~$n(\ce{H^{13}CO^+})$ ratios
used in Models 1 to 3, 4 and 5 respectively. The dashed black line marks the lowest gas temperature in the disk model. }
\end{figure}

The radial abundance profiles for our best-fit models for the \ce{HCO+} and \ce{H^{13}CO+} emission 
shows that isotope-selective processes are occurring in this disk;  
however, the enhancement factors required for \ce{H^{13}CO+} are needed across a significantly larger column density of the disk 
compared to predictions from models, and we need a radial change in the ratio. 
Comprehensive chemical models are required to determine if the \ce{H^{12}CO+}/\ce{H^{13}CO^+} abundance ratio exhibits 
variations  more than a factor of a few relative to the elemental ratio in the outer disk (100's of au)
and how the observed line emission will reflect this.

\subsection{Estimating the gas depletion in the cavity}

We detect \ce{HCO+} emission in the channel maps down to at least 40~au whereas 
the model assumes a gas cavity size of 65~au.
By adding gas to the cavity in Model 3 with an abundance 0.2 times that at the edge of the disk
we can well reproduce the radial profile of the \ce{HCO+} emission. 
The fractional abundance of \ce{HCO+} in the dust cavity reaches a maximum value of  
$6~\times~10^{-9}$. 
Chemical models show that the \ce{HCO+} fractional abundance in the 
inner ($<$ 50~au) regions of full disks can be as high as $10^{-7}$ \citep{2010ApJ...722.1607W, 2015ApJ...807..120A}. 
Hence, if we assume that this value represents an absolute ceiling value for the \ce{HCO+} fractional abundance
then the gas surface density is $\approx$500 times less that in the dust cavity than in the inner edge of the outer disk. 
However, there remain caveats to this estimate related to chemistry.  Because
there is a central dust cavity in this disk then the gas will be less shielded from 
ionising radiation resulting in potentially higher \ce{HCO+} abundances 
than calculated in chemical models of disks without dust cavities. 
\ce{HCO+} is not an optimal tracer of gas mass; hence, the gas depletion in the cavity be should be estimated using CO 
isotopologue observations.

\subsection{A comparison with other Herbig Ae/Be disks}

Molecular line emission from Herbig Ae/Be disks has been shown to trace both sub-structure associated 
with an underlying ringed dust structure and radial structure that arises due to chemistry, such as snowlines 
\citep[e.g., HD 163296, HD~169142:][]{2013ApJ...765...34Q, 2013A&A...557A.132M, 2015ApJ...813..128Q, 2016PhRvL.117y1101I, 2017A&A...600A..72F, 2018A&A...614A.106C}.
With ALMA we now have spatially resolved detections of the \ce{HCO+} and \ce{HCN} 
isotopologues in multiple protoplanetary disks.
Although these molecules are commonly detected in disks around T Tauri 
stars, e.g., IM Lup, AS 209, V4046 Sgr, DM Tau, LkCa 15 \citep[][]{2007A&A...467..163P, 2015ApJ...810..112O, 2015ApJ...814...53G, 2017ApJ...835..231H, 2017ApJ...836...30G},
we will primarily discuss the detections towards HD~97048 in the context of other Herbig Ae/Be disks 
given the differences in chemistry expected between these two disk types \citep[e.g.,][]{2015A&A...582A..88W,2018A&A...616A..19A}.

Both \ce{HCN} and \ce{HCO+} have been detected towards the dust-trap-hosting transition disk 
around the Herbig Ae star, HD~142527 \citep{2013Natur.493..191C, 2014ApJ...782...62R, 2014ApJ...792L..25V, 2015ApJ...811...92C}.  
However, for this source, the molecular emission is strongly influenced by the large cavity (140~au) and the lop-sided dust trap.  
The \ce{HCO+} emission from within the cavity traces non-Keplerian motions that have 
been attributed to radial flows through the gap \citep{2013Natur.493..191C, 2014ApJ...782...62R, 2015ApJ...811...92C},  
and the \ce{HCN} emission is strongly anti-coincident with the dust trap. 
This anti-coincidence is proposed to be due to either the lower dust temperatures in the trap (leading to enhanced freezeout) 
or a high continuum optical depth in the dust trap that is 
suppressing the emergent line emission \citep{2014ApJ...792L..25V}.  
The disk around HD~142527 is a clear example demonstrating the influence that extreme dust structure has 
on the emergent molecular emission.  
Emission from \ce{H^{13}CO+} ($J=8-7$) from HD~142527 has been targeted with ALMA;  
however, it was reported as a non-detection \citep{2015ApJ...812..126C}.

Evidence for multiple rings of \ce{H^{13}CO+} have been detected in the HD~163296 disk and one 
ring of \ce{H^{13}CO+} has been detected in the MWC 480 disk,
but their relation to the underlying dust and/or gas structure 
has not yet been fully investigated using detailed models.
\citep{2017ApJ...835..231H}. 
These observations have a spatial resolution of $\approx$~60 to 70~au.
In both of these sources there is a central cavity in the \ce{H^{13}CO+} emission and
this has been primarily attributed to optically thick continuum emission in the inner disk 
which suppresses the emergent line emission.
For the specific case of \ce{H^{13}CO+}, this may also be partially due to chemistry in the inner disk.
Depending on the location of the \ce{H2O} snowline, which can reside at $\sim 10$s of au from the star in Herbig Ae/Be disks 
\citep[e.g.,][]{2009A&A...501L...5W, 2017ApJ...836..118N}, gaseous \ce{H2O} may be present which facilitates another destruction pathway for \ce{HCO+}. 
Hence, the absence of \ce{H^{13}CO+} emission in the inner regions of protoplanetary disks and protostars has been proposed 
as an indirect indication of the radial location of the water snow line.
This effect has yet to be conclusively demonstrated for a protoplanetary disk but has been 
demonstrated for the protostellar envelope encompassing NGC1333-IRAS2A \citep{2018A&A...613A..29V}.
As yet, there is no evidence for a central cavity strongly depleted in dust in neither HD 163296 nor MWC 480, unlike 
as has been found for HD~97048.
This explains why the \ce{H^{13}CO+} emission profile for HD~97048 is centrally peaked: the emergent emission is 
not suppressed due to the presence of a significant reservoir of mm-sized dust.  

The first detection of \ce{HC^{15}N} in a protoplanetary disk was reported in the MWC 480 disk \citep{2015ApJ...814...53G}.
In this source the \ce{HCN}, \ce{H^{13}CN} and \ce{HC^{15}N} all show similar radial 
distributions which follow the smoothly decreasing dust emission profile. 
From the same study, \ce{HCN} emission from the DM Tau disk, a T Tauri source, shows
a second ring of emission beyond the mm-dust emission at $\sim300$~au. 
The ring is proposed to arise from the result of increased photodissociation of \ce{HCN}  
due to the depletion of micron sized dust interior to the detected molecular ring. 
The \ce{HC^{15}N} emission that we observe appears to follow the radial emission profile of the mm-dust to the spatial resolution of the observations. 
Observations of the main isotopologue \ce{HCN} with improved spatial resolution in the HD~97048 disk are needed confirm the presence of 
rings in this tracer that align with the dust rings.
The \ce{H^{13}CN} emission in the AS 209 disk, another T Tauri disk, is also ringed 
\citep{2017ApJ...836...30G}, and recent ALMA observations have shown clearly that this disk has multiple
rings of dust \citep{2018A&A...610A..24F}. 
This, again, is another example disk with rings of dust and rings of molecular gas. 
\ce{HC^{15}N} has since been detected in the HD~163296 disk but the observations are not of high 
enough S/N to make assertions about the radial distribution in relation to the dust rings \citep{2017ApJ...836...30G}.
Nevertheless, there is growing evidence, including in the analysis presented here, 
that molecular line mission from optically thin tracers may be following, and is thus influenced by, 
the underlying dust distribution in Herbig Ae/Be disks making them interesting test cases.

As stated previously, the disk around Herbig Ae star, HD~100546, was also targeted 
in the same observing campaign during which the data reported here for HD~97048 were collected.
HD 100546 has been shown to also host rings of mm-dust similar to
HD~97048; however, the dust sculpting in HD~100546 is much more extreme with most of the (sub)mm emission 
arising from a narrow ring of dust centred at 26~au with a width of 21~au \citep{2014ApJ...791L...6W, 2019ApJ...871...48P}. 
This has been proposed to arise due to a combination of radial drift and the sequential formation of 
two giant planets in the disk \citep{2015A&A...580A.105P}.
We do not detect neither
\ce{H^{13}CO+} nor \ce{HC^{15}N} in the disk around HD~100546.
To date, there are no reported detections of HCN neither
such that it is not currently possible to draw any concrete conclusions from the non-detections reported here 
for this species.  
\citet{2015MNRAS.453..414W} report the detection of \ce{HCO+} $J=1-0$ with ATCA towards HD~100546, so the non-detection of 
of the higher energy $J=4-3$ transition of the isotopologue, \ce{H^{13}CO+}, is perhaps surprising, given that 
emission from the main isotopologue is expected to be optically thick. 
However, we have detected \ce{SO} in this disk, proposed to be tracing either a disk wind or a circumplanetary
disk \citep{2018A&A...611A..16B}.  
In contrast, we do not detect \ce{SO} in the HD~97048 disk.

The differences between the molecular content of HD~97048 and HD~100546 may be explained 
by them being in different stages of evolution where
HD~97048 is still forming planets whereas the HD~100546 is not.    
In support of this explanation, the same direct-imaging techniques have been used to search for the presence of planets in both sources; however, whilst
there is a confirmed planet candidate in the HD~100546 disk, there are none yet confirmed to reside in the HD~97048 disk using direct imaging methods 
\citep[e.g.,][]{2012A&A...538A..92Q, 2013ApJ...766L...1Q}. 
However, note the recent indirect detection of a few Jupiter mass planet at 130~au via kinematic analyses of the CO line emission \citep{2019arXiv190702538P}.
This is a significantly lower mass planet than detected in HD~100546 that also appears to be undergoing accretion, which supports this hypothesis.

\subsection{What is the effect of the ringed dust (and gas) structure on the chemistry?}

We find that a radially 
smooth profile cannot fit the \ce{H^{13}CO+} observations and the \ce{H^{12}CO+}/\ce{H^{13}CO^+} abundance ratio
across the disk. 
As this is not wholly consistent with predictions from chemical fractionation models we now explore the possible effect 
of the dust rings on the disk chemistry.

From ALMA observations we know that the mm-sized dust is distributed in three rings \citep{2016ApJ...831..200W, 2017A&A...597A..32V}.
Scattered light observations with SPHERE show that at the
disk surface there are four rings, two of which are located at comparable positions to the mm-sized dust rings \citep{2016A&A...595A.112G}.
Due to the matching rings in both the mm-sized and micron-sized grains 
planet-disk interactions have been proposed as the best explanation for the 
dust rings in the HD~97048 disk \citep{2016ApJ...831..200W, 2016A&A...595A.112G, 2017A&A...597A..32V}.
From the non-detections of direct emission from planets in the gaps and by comparing the observed rings to 
models, limits can be placed on the masses of the anticipated planets.
\citet{2017A&A...597A..32V} estimated the mass of the inner-most planet, located between 2.5 and 11~au, to be $\approx$~0.7~\ce{M_{J}} 
guided by models from \citet{2016MNRAS.459.2790R}.
Low mass planets, $\approx$~1~\ce{M_{J}}, will create a narrower and shallower cavity in the gas
surface density than in the mm-sized dust which forms more distinct rings 
\citep{2012ApJ...755....6Z, 2013A&A...560A.111D, 2016MNRAS.459L..85D}.
\citet{2019arXiv190702538P} indirectly detect a planet of a few Jupiter masses at 130~au, which is consistent with upper-limits from 
VLT/SPHERE \citep{2016A&A...595A.112G}.
Hence, it is possible that perturbations in the gas surface density are sufficiently shallow that they 
are not visible in our \ce{CO} and \ce{HCO+} isotopologue observations, or conversely, that the gas gaps are deep, but also sufficiently narrow 
that subsequent chemical inhomogeneities are resolved out.

We explore here whether or not variations in the underlying gas density due to forming planets 
may be responsible for the observed radial variation in the \ce{H^{12}CO+}/\ce{H^{13}CO+}
abundance ratio across the disk. This is contrast to the scenario discussed in Section 6.1 (see Figure 9a). Although our data do not have the resolution to resolve
the proposed gas gaps, the presence of such gaps may be revealed through analysing global
abundance ratios, that average over the chemical sub-structure. 
A drop in the column density of small dust grains at the location 
of the millimetre-sized dust gaps will provide less shielding from photodissociating {\em and} photoionising radiation. 
This will increase the ionisation fraction within the dust gaps relative to the dust rings.  
Given that \ce{HCO+} is the dominant cation in the molecular layer, its abundance may also increase in the dust 
gaps and this may mask the presence of gas depletion therein because the overall column density of \ce{HCO+} may not 
change significantly. In this case the difference between the \ce{HCO+} and \ce{H^{13}CO+} abundance distributions would only 
be due to isotope-selective chemistry (Figure 9b). Also, 
because of the non-linear nature of the chemistry, a reduction in gas density can have a profound effect on the abundances of 
more minor species because the rates of two-body ion-molecule reactions are proportional to $n^{2}$.  
Hence, it is possible that this can amplify the effects of isotope-selective photodissociation by e.g., slowing down the 
reformation of \ce{^{13}CO} relative to \ce{^{12}CO}.  
This would reduce the abundance of \ce{H^{13}CO+} relative to \ce{HCO+} in the disk, in contrast to our observations (Figure 9c).

An additional explanation to consider is the effect of the dust structure on the gas temperature. 
Models show that 
the gas temperature relative to the dust temperature can decrease in the disk atmosphere due to grain growth and settling and,  
at the location of the dust gaps \citep{2017A&A...605A..16F, 2018ApJ...867L..14V}.
A decrease in gas temperature could be traced by isotope fractionation (see Figure 9d) but
this remains speculation until dedicated physico-chemical models are available 
that include isotope-selective chemistry as well as ringed dust structure.

\begin{figure*}
\includegraphics[trim={0 0 0 0},clip,width=\hsize]{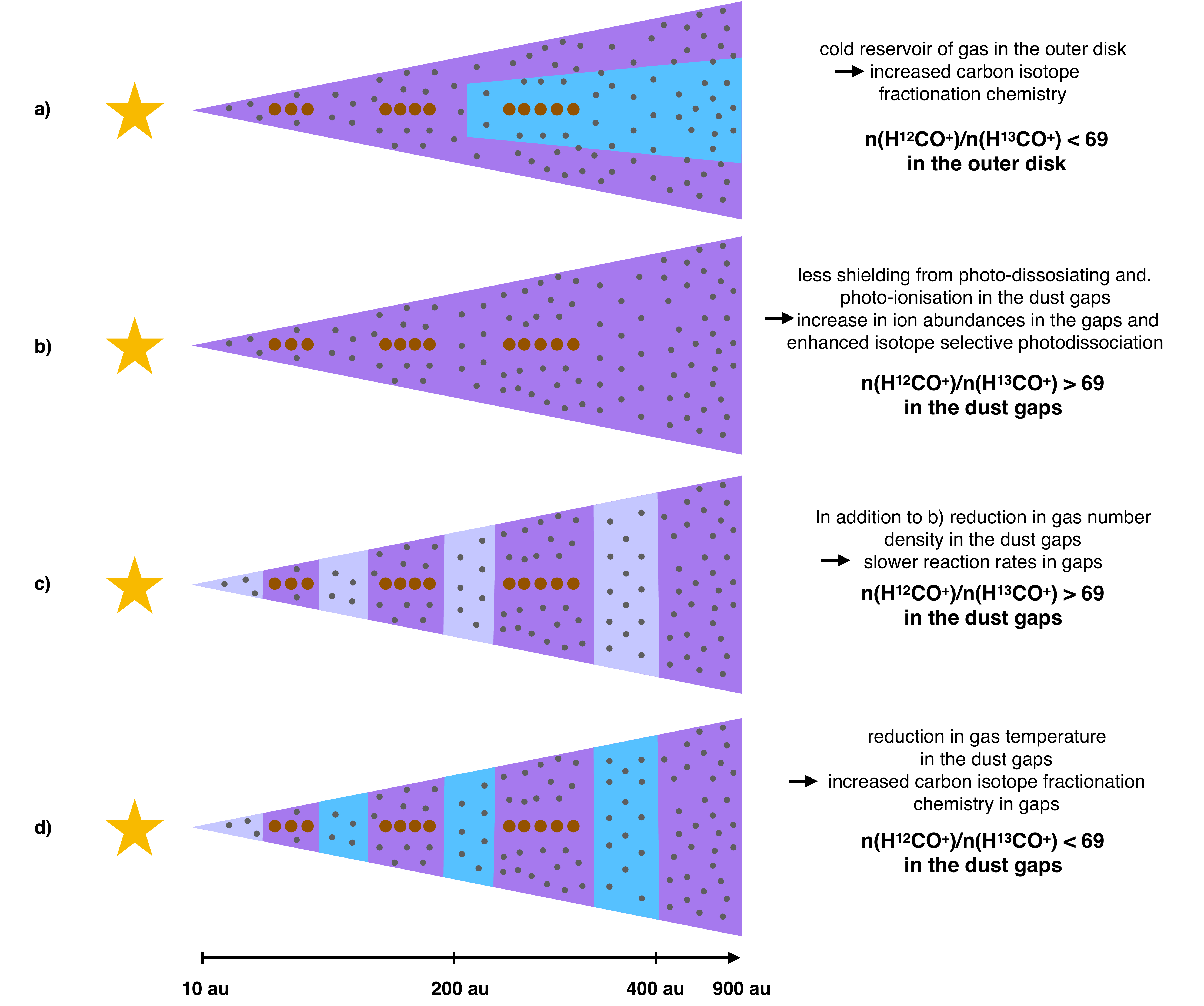}
\caption{Cartoon illustrating the possible gas and dust distributions in the HD~97048 disk.
The mm-sized dust is represented by brown circles, the micron-sized dust is represented by grey circles, the 
gas and depleted gas are represented by purple and lilac respectively and the light blue corresponds to a cold gas reservoir.}
\end{figure*}

\section{Conclusion}

The observations of disks around Herbig Ae/Be stars with ALMA so far have shown the incredible potential of these 
sources as laboratories to study giant planet formation.  
With ALMA, we have shown for the first time that the disk molecules \ce{H^{13}CO+} and \ce{HC^{15}N},
are may be tracing sub-structure in the HD~97048 protoplanetary disk.
Our favoured by-eye fit to the \ce{H^{13}CO+} observations require a model with an enhancement in fractional abundance relative to the canonical 
isotope ratio in the outer disk.  
Since the \ce{HCO+} models do not require such enhancement factors, chemical fractionation 
is the simplest explanation for the emergent line emission.
The enhancement in \ce{H^{13}CO+} relative to \ce{HCO+} can be explained by a cold reservoir of gas in the outer disk ($\lesssim$~10~K, $\gtrsim$~200~au)
Through a comparison with disk models that include 
carbon fractionation we argue that isotope selective chemistry cannot solely explain the observations and we discuss
the effects of the ringed dust, and possible ringed gas, on the disk thermal and chemical structure 
on the interpretation of observations.   

Gas giant planets on wide orbits are the potential exoplanet population currently probed
with ALMA via the ringed depletion of continuum emission. 
As disks around Herbig Ae stars are the progenitors to multiple 
gas giant systems \citep[e.g. HR 8799][]{2008Sci...322.1348M} studying these disks is of great importance. 
These new detections further motivate and highlight the need for 
future studies on the distribution of molecular gas in these planet-forming disks.
Future ALMA observations of HD~97048 need to have the spatial resolution to properly resolve the ringed structure in the dust and in the different gas tracers, and  
will allow for the determination of the size of central cavity in different tracers.
This will enable a detailed study on the dominant chemical processes in the disk and allow us to make clearer distinctions between the different scenarios we have presented. 
For further progress in this field 
disk models that accurately represent the dust and gas distribution in disks are
necessary. Specifically, models which take into account the gas within the dust cavity and the multiple rings of gas and dust depletion.
Such models will maximise the power of molecular line emission to provide much needed information 
on the physical and chemical conditions in planet-forming disks.

\begin{acknowledgements}
We thank the referee for their constructive report. 
This paper makes use of the following ALMA data: ADS/JAO.ALMA$\#$2011.0.00863.S
and 2012.1.00031.S. ALMA is a partnership of ESO 
(representing its member states), NSF (USA) and NINS (Japan), together with NRC (Canada), NSC and ASIAA 
(Taiwan), and KASI (Republic of Korea), in co-operation with the Republic of Chile. 
The Joint ALMA Observatory is operated by ESO, AUI/NRAO and NAOJ. 
A.B. acknowledges the studentship 
funded by the Science and Technology Facilities Council of the United Kingdom (STFC). 
C.W. acknowledges funds from the University of Leeds.  
J.D.I. and C.W. acknowledge support from the STFC under ST/R000549/1.
\end{acknowledgements}

\bibliography{hd97048_accepted.bib}

\begin{appendix}
\onecolumn

\newpage
\section{\ce{H^{13}CO^+} and \ce{HC^{15}N channel maps}}
\begin{figure*}[h!]
\includegraphics[trim={1cm 1cm 0 4cm},clip,width=\hsize]{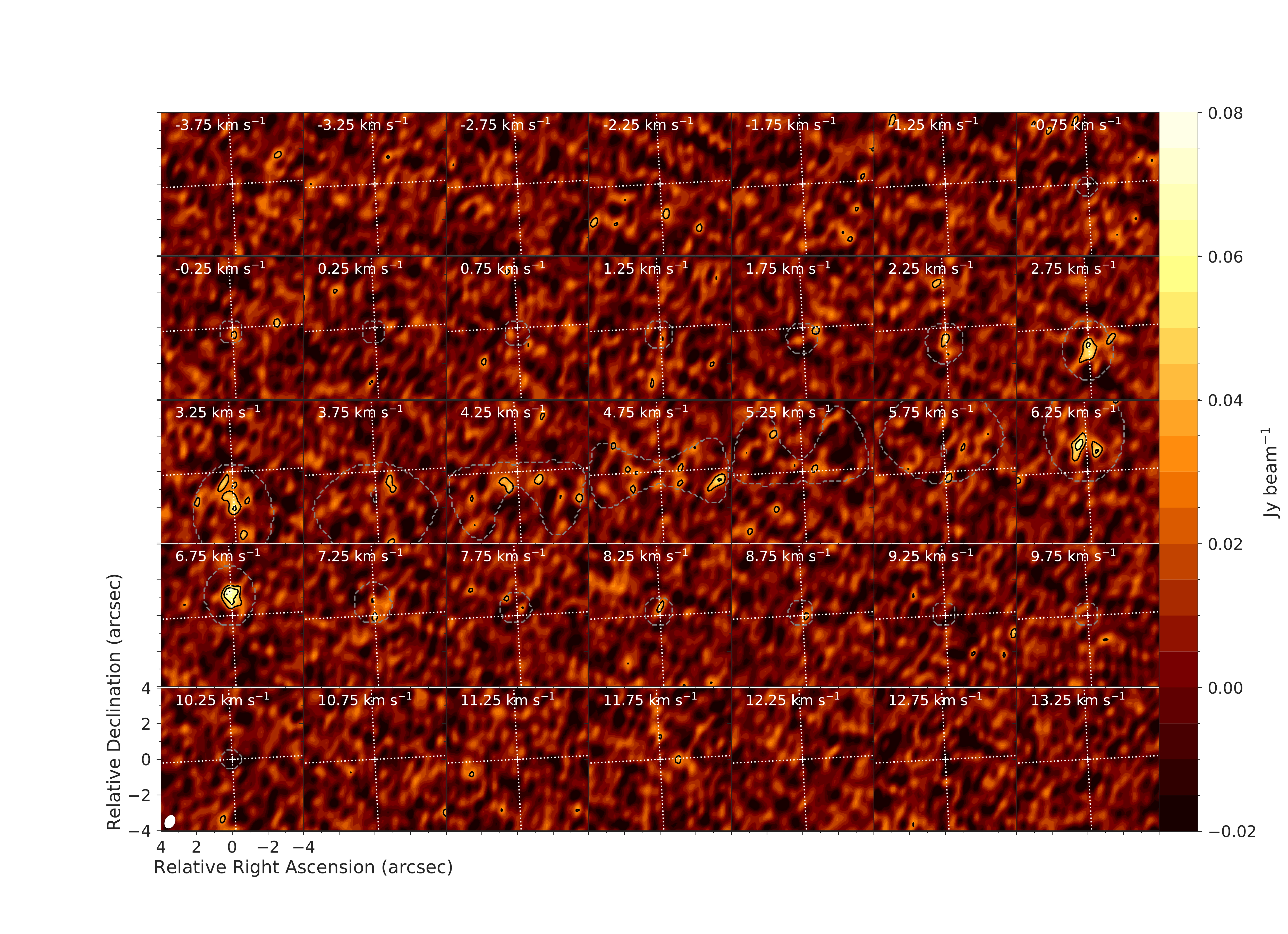}
\includegraphics[trim={1cm 1cm 0 4cm},clip,width=\hsize]{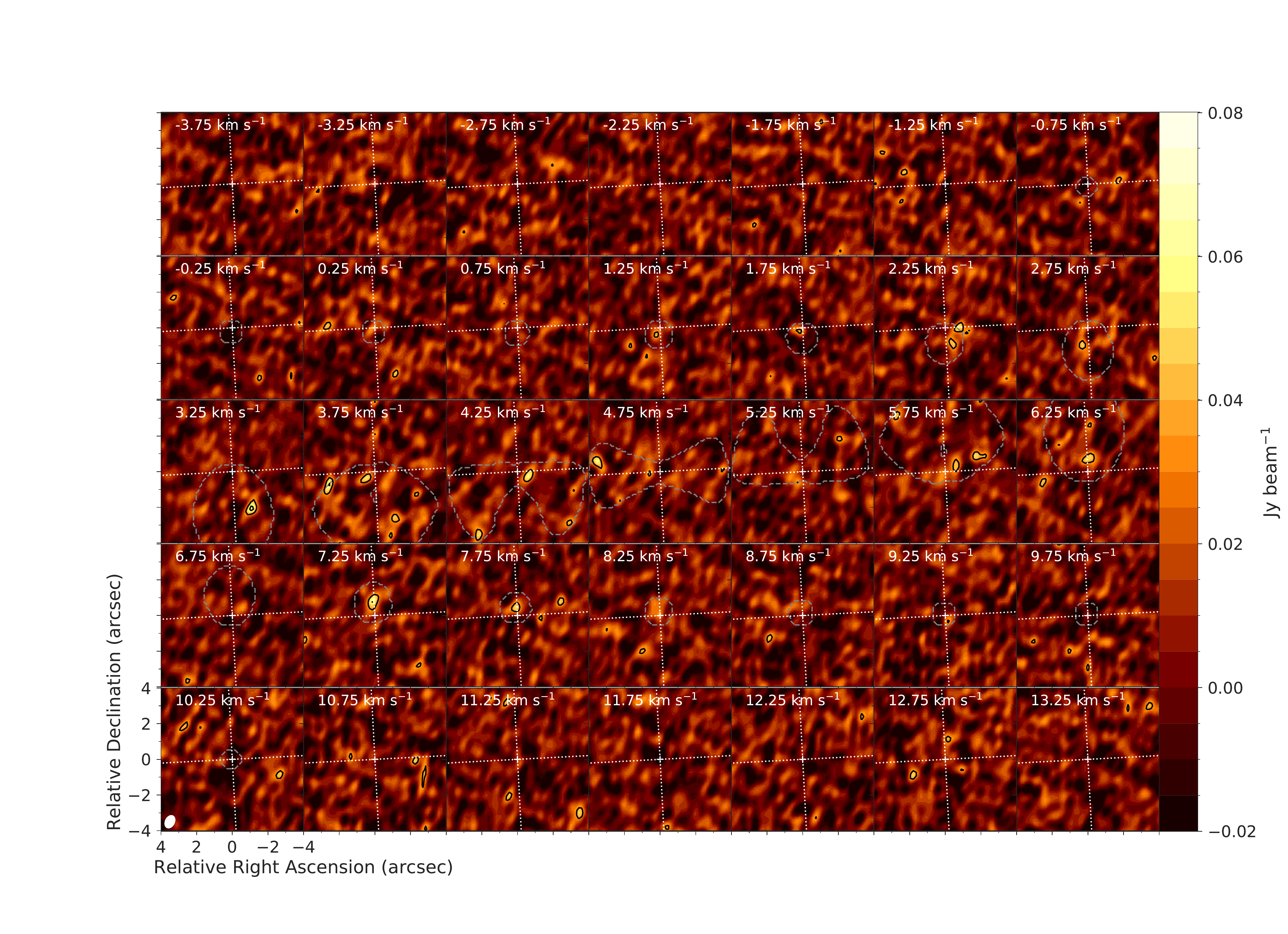}
\caption{\ce{H^{13}CO^{+}}(top) and \ce{HC^{15}N} (bottom) channel maps with a velocity resolution of 0.5~km~s$^{-1}$. 
The r.m.s. for each map are 0.011 Jy~beam$^{-1}$~km~s$^{-1}$ channel$^{-1}$ and 0.012 Jy~beam$^{-1}$~km~s$^{-1}$ channel$^{-1}$ respectively. 
Emission is detected in each map with a S/N of 6.8 and 5.7 respectively.
The black contours are at the 3 and 5$\sigma$ level.
The grey dashed contour shows the Keplerian mask used and
the dotted white lines mark the major and minor axes of the disk.}
\label{figureb1}
\end{figure*}

\section{\ce{HCO^+} and \ce{H^{13}CO^+} residual channel maps}
\begin{figure*}
\centering
\includegraphics[trim={1cm 1cm 0 2cm},clip,width=\hsize]{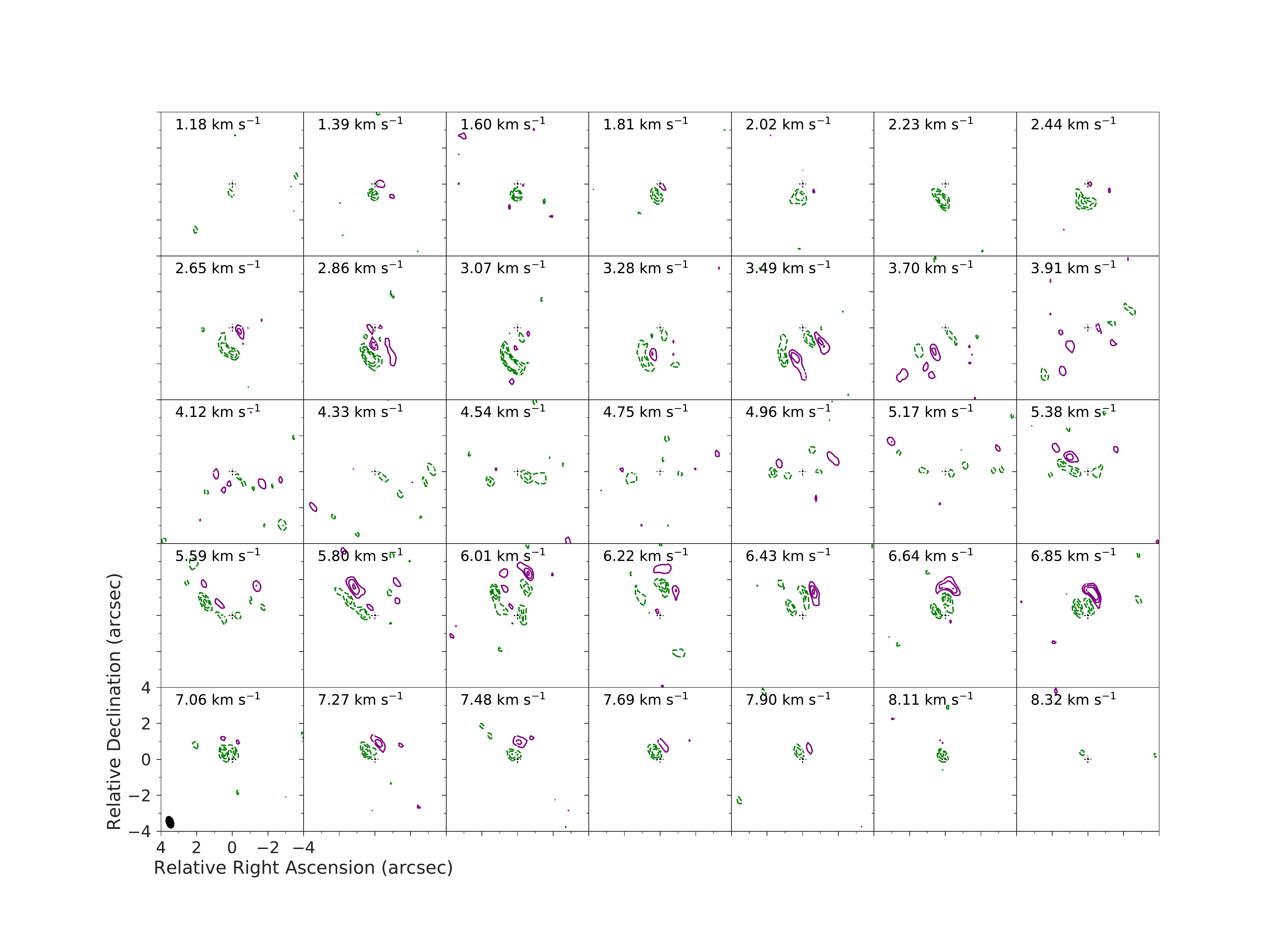}
\includegraphics[trim={1cm 1cm 0 2cm},clip,width=\hsize]{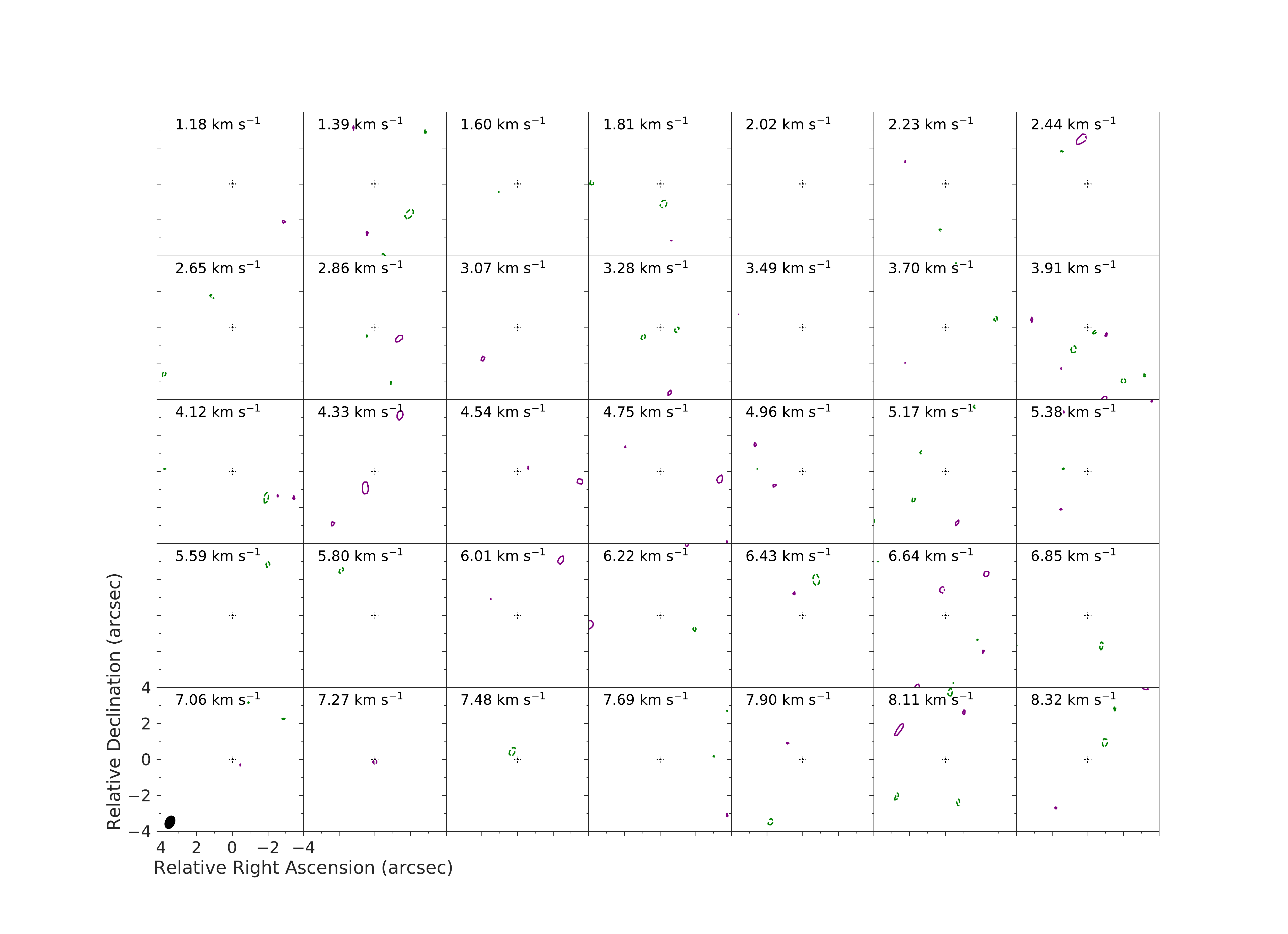}
\caption{Residual (data - model) \ce{HCO+} (top) and \ce{H^{13}CO+} (bottom) channel maps at a velocity resolution of 0.21~km~s$^{-1}$. 
The r.m.s. is 0.016 Jy~beam$^{-1}$~km~s$^{-1}$ channel$^{-1}$ 
the purple and green contours are $\pm~3,5,7\sigma$ respectively.}
\label{figure8}
\end{figure*}

\end{appendix}

\end{document}